\PassOptionsToPackage{bookmarks=false}{hyperref}
\documentclass[sigconf,screen]{acmart}

\usepackage{algorithmic}
\usepackage{graphicx}
\usepackage{textcomp}
\usepackage{xcolor}

\usepackage{makecell}
\usepackage{tabularx}
\usepackage[T1]{fontenc}
\usepackage{comment}
\usepackage{xcolor}
\usepackage{multirow}
\usepackage{array}
\usepackage{soul}
\usepackage{threeparttable}
\usepackage{balance}
\usepackage{diagbox}
\usepackage{ulem}
\usepackage{subfigure}
\usepackage{microtype} 
\usepackage{calc}

\usepackage{enumitem}

\usepackage{url}

\usepackage{float}
\usepackage{stfloats}
\usepackage{xspace}

\newcommand\SO{Stack Overflow\xspace}
\newcommand{\app}{KGE4AR\xspace}
\newcommand{\parabf}[1]{\noindent\textbf{#1}}

\newcommand\update[1]{{\color{red}#1}}

\newcommand\yiling[1]{{\color{red} \textbf{yiling: }\color{blue}\color{blue}#1}}

\newcommand\etal{\textit{et al. }}

\newcommand\ie{\textit{i.e., }}
\newcommand\eg{\textit{e.g., }}

\newcommand\newRAPIM{RAPIM$\ast$}
\newcommand\newDAPIMap{D2APIMap$\ast$}

\newcommand{\smalltexttt}[1]{{\small \texttt{#1}}}

\AtBeginDocument{%
	\providecommand\BibTeX{{%
			\normalfont B\kern-0.5em{\scshape i\kern-0.25em b}\kern-0.8em\TeX}}}

\setcopyright{acmlicensed}
\acmPrice{15.00}
\acmDOI{10.1145/3611643.3616305}
\acmYear{2023}
\copyrightyear{2023}
\acmSubmissionID{fse23main-p579-p}
\acmISBN{979-8-4007-0327-0/23/12}
\acmConference[ESEC/FSE '23]{Proceedings of the 31st ACM Joint European Software Engineering Conference and Symposium on the Foundations of Software Engineering}{December 3--9, 2023}{San Francisco, CA, USA}
\acmBooktitle{Proceedings of the 31st ACM Joint European Software Engineering Conference and Symposium on the Foundations of Software Engineering (ESEC/FSE '23), December 3--9, 2023, San Francisco, CA, USA}
\received{2023-02-02}
\received[accepted]{2023-07-27}

\AtBeginDocument{%
	\providecommand\BibTeX{{%
			\normalfont B\kern-0.5em{\scshape i\kern-0.25em b}\kern-0.8em\TeX}}}
\begin{document}


\title{Recommending Analogical APIs via Knowledge Graph Embedding}

\renewcommand{\shorttitle}{Recommending Analogical APIs via Knowledge Graph Embedding}

\author{Mingwei Liu}\authornote{M. Liu, Y. Yang, Y. Lou, X. Peng, Z. Zhou, X. Du, and T. Yang are with the School of Computer Science and Shanghai Key Laboratory of Data Science, Fudan University, China.}
\affiliation{%
  \institution{Fudan University}
  \country{China}
}

\author{Yanjun Yang}\authornotemark[1]
\affiliation{%
  \institution{Fudan University}
  \country{China}
}

\author{Yiling Lou}\authornotemark[1]
\authornote{Y. Lou is the corresponding author (email: yilinglou@fudan.edu.cn).}
\affiliation{%
  \institution{Fudan University}
  \country{China}
}

\author{Xin Peng}\authornotemark[1]
\affiliation{%
  \institution{Fudan University}
  \country{China}
}

\author{Zhong Zhou}\authornotemark[1]
\affiliation{%
  \institution{Fudan University}
  \country{China}
}

\author{Xueying Du}\authornotemark[1]
\affiliation{%
  \institution{Fudan University}
  \country{China}
}

\author{Tianyong Yang}\authornotemark[1]
\affiliation{%
  \institution{Fudan University}
  \country{China}
}


\begin{abstract}
	Library migration, which replaces the current library with a different one to retain the same software behavior, is common in software evolution. An essential part of this is finding an analogous API for the desired functionality. However, due to the multitude of libraries/APIs, manually finding such an API is time-consuming and error-prone. Researchers created automated analogical API recommendation techniques, notably documentation-based methods. Despite potential, these methods have limitations, e.g., incomplete semantic understanding in documentation and scalability issues.

In this study, we present \app, a novel documentation-based approach using knowledge graph (KG) embedding for recommending analogical APIs during library migration. \app introduces a unified API KG to comprehensively represent documentation knowledge, capturing high-level semantics. It further embeds this unified API KG into vectors for efficient, scalable similarity calculation. We assess \app with 35,773 Java libraries in two scenarios, with and without target libraries. \app notably outperforms state-of-the-art techniques (e.g., 47.1\%-143.0\% and 11.7\%-80.6\% MRR improvements), showcasing scalability with growing library counts.

\end{abstract}


\maketitle

\section{Introduction}
\label{sec:introduction}
Third-party libraries are pivotal in modern software development, enhancing quality and productivity~\cite{softw94libreuse,ese07libreuse,fse12libmigration,ese18libmigration}. However, as software and libraries evolve rapidly, current libraries might turn unsuitable due to factors like sustainability failures~\cite{fse17projectfail,fse18pypi}, license restrictions~\cite{softw12license,ase14license}, lacking features~\cite{saner21libmigration}, and security/performance issues~\cite{saner21libmigration}. This necessitates \textit{library migration}, where developers replace current libraries with new ones to re-implement the same software behavior. Such migrations are common in software evolution~\cite{fse21libmigration}; for example, He \etal~\cite{fse21libmigration} found 8.98\% \textasciitilde{} 28.72\% of 17,426 open-source projects underwent at least one library migration.

\textcolor{black}{However, library migration~\cite{fse12libmigration,ese18libmigration,sevis19libevolution,CASCON18funmap,icpc19apimap} is a very time-consuming, labor-intensive, and error-prone task for developers in practice. For example, the prior study shows that some developers even spend up to 42 days for library migration~\cite{icpc19apimap}. Given the currently-used library (called \textbf{source library}) and the API (called \textbf{source API}), one essential part in library migration is to find an analogical library (called \textbf{target library}) and an analogical API (called \textbf{target API}), which can provide the same functionality as current ones. However, manually finding analogical API is a heavy burden for developers, since they need to read length API documentation and code snippets of potentially-analogical APIs~\cite{icpc19apimap, icsme19apimig, CASCON18funmap, asc20apimigration} while there are an extremely large number and fast changes of third-party libraries and APIs (e.g., as of January 2020 there are 35,773 common Java libraries with 15,441,057 APIs on Libraries.io~\cite{libariesiodata}).} 

To reduce efforts in \textcolor{black}{manually searching and reading API documentation and code snippets for determining the analogical relationship between libraries and APIs}, many techniques have been proposed to recommend suitable target libraries or analogical target APIs. In this work, we focus on analogical API recommendation.
Researchers have harnessed diverse resources to facilitate such recommendations~\cite{icsme19apimig,CASCON18funmap,asc20apimigration,icpc19apimap}, including evolution history~\cite{wcre13funcmap,ase14apimap}, online Q\&A interactions~\cite{tse19apimigration}, and API documentation~\cite{asc20apimigration,icpc19apimap,scam15apimap,tse19apimigration,jsep17tmap,ksem17apimap,icpc20apimigration}.
Among these, documentation-based API recommendation has been intensively studied in the literature, since API documentation is prevalent and at low cost to collect while other information could be time-consuming to collect and are not always available.  
For a given source API, existing documentation-based techniques calculate the textual similarity between each candidate API and the source API (\eg the textual similarity between two API functionality descriptions in the documentation), and then recommend the candidate API with the highest similarity as the target API. 

While promising, current documentation-based API recommendation techniques face two limitations.
First, their way of calculating textual similarity falls short in capturing semantic-level connections in API documentation. 
These techniques mainly calculate the textual similarity based on the overlapping tokens~\cite{jsep17tmap} or measure the token similarity without contextual consideration~\cite{icpc20apimigration}.
This can lead to identifying analogical semantics in API descriptions that share similar noun phrases but differing action verbs (\eg ``set S3 Object content'' vs. ``get S3 Object content length'').
Additionally, these techniques seldom consider domain knowledge when calculating textual similarity. For example, JSON arrays, JSON objects, keys, and values are all JSON-related concepts that often occur in APIs related to JSON processing.
Concepts, in the context of our work, refer to domain-specific entities or terms, often represented by noun phrases, that capture specific elements or ideas within the API domain.
Without considering such conceptual relationships, the estimation of semantic similarity/relevance between two analogical APIs might be underestimated.
Second, these techniques typically compute similarity pairwise, posing computational challenges with a vast number of candidate APIs. For example, envision a library like TestNG~\cite{testng}, encompassing over 4,000 candidate APIs. Existing techniques require performing over 4,000 pairwise comparisons to calculate the similarity between a single source API and all the candidate APIs. 
This exhaustive calculation demands substantial online costs and becomes prohibitively expensive when multiple target libraries are involved.

To address this, we propose \textbf{\app}, a novel documentation-based method leveraging \ul{\textbf{K}}nowledge \ul{\textbf{G}}raph \ul{\textbf{E}}mbedding \ul{\textbf{for}} analogical \ul{\textbf{A}}PI \ul{\textbf{R}}ecommendation \textit{effectively and scalably}. \app constructs a unified API knowledge graph (KG) for third-party libraries from API documentation, leveraging graph embedding to represent nodes and edges as numeric vectors. It efficiently retrieves the most similar API for a given source API from the embedded KG. Compared to previous approaches, \app introduces two technical innovations. Firstly, it presents a novel \textit{unified API KG} that comprehensively represents three types of documentation knowledge across diverse libraries, better capturing overall semantics in API documentation. Secondly, \app proposes \textit{embedding the unified API KG}, enhancing efficiency and scalability by streamlining analogous API vector retrieval via vector indexing.

To implement \app, we build a unified API KG consisting of 59,155,631 API elements sourced from 35,773 Java libraries. This KG comprises a total of 72,242,099 entities and 289,122,265 relations connecting these entities.
We evaluate \app in two API recommendation scenarios: with and without target libraries. When given the target libraries, \app achieves 47.1\%-143.0\% and 41.4\%-95.4\% improvements over the baselines in terms of MRR and Hit@10, respectively; while without a given target library, \app substantially outperforms existing analogical API recommendation techniques by achieving 11.7\%-80.6\%, 26.2\%-72.0\%, and 33.2\%-116.5\% improvements in terms of MRR, precision, and recall, respectively.
We also evaluate the scalability of \app and find that it scales well with an increasing number of libraries. Furthermore, we extensively investigate the impact of different design choices in \app.

In summary, this work makes the following contributions:
\begin{itemize}[itemsep=2pt,topsep=0pt,parsep=0pt, leftmargin=20pt]
    \item \textbf{Novel Approach:} 
    We introduce \app, a documentation-based analogical API recommendation method that builds a unified API KG for numerous libraries, offering scalable recommendations via KG embedding.
    \item \textbf{Thorough Evaluation: } 
    We thoroughly evaluate \app{} through effectiveness comparisons in two API recommendation scenarios, scalability assessment across various library quantities, and analysis of design choice implications.
    \item 
    \textbf{Public Benchmark}: We release a benchmark for extensive analogical API evaluations across numerous libraries.
\end{itemize}


\section{Background and Related Work}
In this section, we discuss related work in analogical API recommendation and knowledge graphs in software engineering.

\subsection{Analogical API Recommendation}
Existing analogical API recommendation techniques leverage various sources like evolution history~\cite{icsme19apimig,ase14apimap}, online posts~\cite{tse21developerneed}, and API documentation~\cite{asc20apimigration,icpc19apimap,scam15apimap,tse19apimigration,jsep17tmap,ksem17apimap, icpc20apimigration} to find suitable target APIs. Evolution-history-based methods~\cite{wcre13funcmap} use evolution history (\eg code changes) to mine frequently co-occurring API pairs, while documentation-based ones~\cite{jsep17tmap,asc20apimigration,icpc19apimap,scam15apimap,tse19apimigration,ksem17apimap} calculate textual similarity using API-related text (\eg descriptions). We concentrate on documentation-based recommendation due to its prevalence, low cost of data collection, and recent research emphasis.

Existing documentation-based API analogical techniques mainly fall into two categories, \eg supervised learning based~\cite{asc20apimigration} and unsupervised learning based  ones~\cite{scam15apimap,jsep17tmap,icpc19apimap,tse19apimigration,ksem17apimap,icpc20apimigration,saner16libmigration}.
For supervised learning-based techniques,  Alrubaye \etal~\cite{asc20apimigration} propose to train a machine learning model (\ie boosted decision tree) for analogical API inference based on the features extracted from API documentation (\eg the similarity of their method descriptions, return type descriptions, method names, and class names) and leverage the trained model to predict the probability of an unseen API pair being analogical. 
Different from supervised techniques that require a large amount of labeled data, unsupervised learning-based techniques often vectorize APIs in an unsupervised way and then recommend analogical APIs based on vector similarity. 
For example, Zhang \etal \cite{icpc20apimigration} leverage the Word2Vec model to vectorize the API functionality description, API parameters, and API return values, and then calculate a joint similarity based on these vectors.

Although achieving promising effectiveness, existing document\-ation-based techniques suffer from two major drawbacks. First, they calculate the textual similarity based on the overlapping tokens~\cite{jsep17tmap} or measure the token similarity without considering the whole context~\cite{icpc20apimigration}, thus cannot well capture the semantic-level similarity in API documentation. Second, they calculate the pair-wise similarity between all APIs in an exhaustive way, thus suffering from the scalability issue when the number of APIs is large. 
To address these issues, our work makes the first attempt to comprehensively and structurally represent the knowledge in API documentation with a novel \textit{unified API KG}. 
In addition, we further leverage the KG embed to enable more effective and scalable similarity calculation. 
Our evaluation results also demonstrate our improvements over existing documentation-based techniques.



\subsection{Knowledge Graph in Software Engineering}
In the domain of software engineering, researchers have established knowledge graphs for diverse objectives, encompassing API concepts~\cite{fse19apisummary,jos2021automatic}, API caveats~\cite{icsme2018apicaveat}, API comparison~\cite{ase20apicomp}, API documentation~\cite{icsme18docgen,icsme20docgen}, domain terminology~\cite{fse2019glossary,wang2023xcos,TSE23CONCEPTLINK}, programming tasks~\cite{fse22taskkg}, ML/DL models~\cite{tse2023aitaskmodelkg}, and bugs~\cite{wang2017construct,su2021reducing}. 
Our work applies the API knowledge graph in a task that is distinct from existing work, namely analogical API recommendation.
In addition, since targeting different tasks, the design and focus of our API knowledge graph are also different from existing ones. For example, the existing API knowledge graph constructed for API misuse detection~\cite{ase2020apimiuse} mainly includes the call-order and condition-checking relations between APIs, while our API knowledge graph focuses on three types of knowledge (\ie API structures, API functionality descriptions, and API conceptual relationships) in API documentation which are helpful for analogical API recommendation. Moreover, we also propose a novel knowledge graph embedding to enable more effective and more scalable analogical API recommendation.

\subsection{Knowledge Graph Embedding}
Knowledge graph embedding (KGE) uses low-dimensional vectors to represent entities and relationships in a knowledge graph, capturing semantic relationships between entities~\cite{kgesurvey}. KGE models map entities into a vector space, where similar ones are closer. They excel in applications like question answering, recommendations, and knowledge graph completion~\cite{kgesurvey,kglsurvey}. Common KGE approaches are TransE, TransR, and DistMult~\cite{transE,transr,DistMult}. These methods encode KG triples (head entity, relation, tail entity) into continuous vector representations. For instance, TransE treats entities and relations as vectors, defining relationships as translations from head to tail entities~\cite{transE}. We employ KGE to embed a unified API KG for analogical API recommendation.

\section{Approach}
\label{sec:approach}
\begin{figure}
	\centering
	\vspace{-2mm}
	\includegraphics[width=0.87\columnwidth]{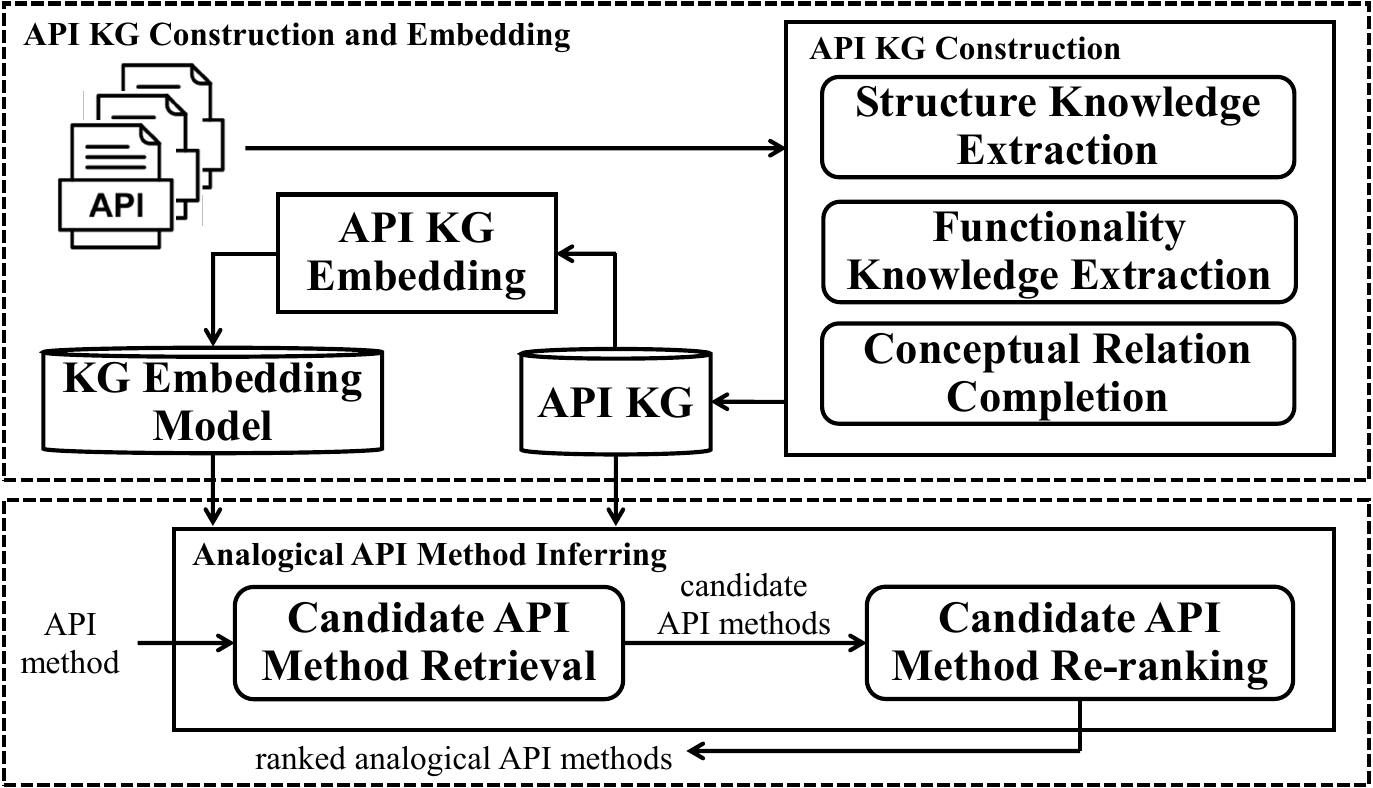}
	\vspace{-4mm}
	\caption{\textcolor{black}{Overview of \app}}
	\vspace{-7mm}
	\label{fig:overview}
\end{figure}


As shown in Figure~\ref{fig:overview}, \app includes three phases, \ie API KG construction, API KG embedding, and analogical API method inferring. Given the API documentation from a large number of libraries as inputs,  \app{} first constructs a unified API KG (Section~\ref{sec:kg_construct}) and then trains an embedding model to embed the constructed KG (Section~\ref{sec:kg_embed}). Lastly, for a given source API, \app{} returns its analogical API based on the embedded KG (Section~\ref{sec:app_infer}). Note that the first two phases only need to be run once. Once the unified KG is constructed and embedded, \app{} \textcolor{black}{can} recommend analogical APIs for the given API efficiently.  In particular, \app mainly has two technical novelties. 

\textbf{Novelty 1: a unified API KG for a large number of libraries.} We propose constructing a unified API KG for a substantial library count (\eg 35,773 Java libraries in this study). Our API KG comprises three knowledge types found in documentation, which often resemble analogical APIs: (1) \textit{API structures} (e.g., package structures, class definitions, method declarations), (2) \textit{API functionality descriptions} (e.g., \textit{``get the number of elements in the JSONArray''}), and (3) \textit{API conceptual relationships} (i.e., API concepts and their relationships like \textit{``belong to''}).
Unlike existing approaches that focus solely on API structures or functionality descriptions presented as token sequences, our unified API KG offers a broader, structural representation encompassing all three knowledge types. This includes a novel category—API conceptual relationships—previously unexplored. A graphical structure inherently suits the structured unification of multi-type data, thus effectively capturing the higher-level semantics within API documentation.
 

\textbf{Novelty 2: a KG embedding-based similarity calculation.}
We propose embedding the unified API KG, representing each KG API as a vector. KG embedding offers two advantages. First, it effectively preserves structural and semantic data in the unified KG. Second, it expedites similarity calculations between APIs in the KG. Retrieving similar API vectors from a database via vector indexing is highly efficient. Unlike existing methods requiring exhaustive similarity calculations for all API pairs, our KG embedding enables a more efficient and effective approach to similarity calculation.

\subsection{API Knowledge Graph Construction}
\label{sec:kg_construct}
In this phase, \app constructs a unified API KG for a large number of libraries based on their API documentation. The API KG construction mainly consists of three steps. (1) Structure knowledge extraction: \app first extracts all API elements (\eg packages, classes/interfaces, methods, fields, parameters) and their relationships from the documentation to form a basic skeleton of the API KG; (2) Functionality knowledge extraction: \app then extracts the functionality knowledge of the API libraries, \ie the standardized functionality expressions of the methods (including functionality verbs, functionality categories, and phrase patterns) and the involved concepts, from the names and text descriptions of methods; (3) Conceptual relation completion:  \app completes conceptual relations between API elements and concepts by analyzing names and text descriptions of API elements and concepts. In this way, API elements from different libraries can be related to each other based on shared type references (\eg types of method parameters and return values), functionality expressions, and concepts.



\subsubsection{Schema of the Unified API Knowledge Graph}
\label{sec:kg_schema}
\begin{figure}
	\centering
	\vspace{-2mm}
	\includegraphics[width=0.9\columnwidth]{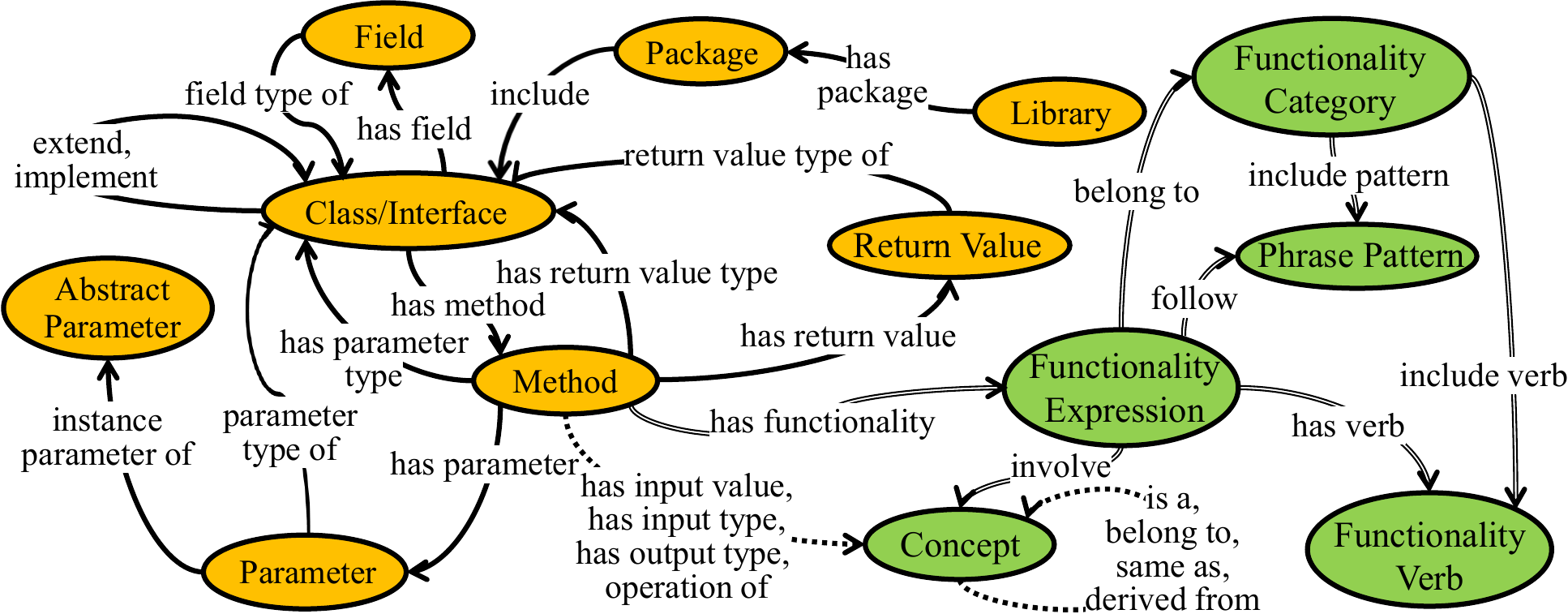}
	\vspace{-4mm}
	\caption{Schema of API KG}
	\label{fig:kgschema}
\end{figure}

\begin{figure}
	\centering
	\vspace{-2mm}
	\includegraphics[width=1.0\columnwidth]{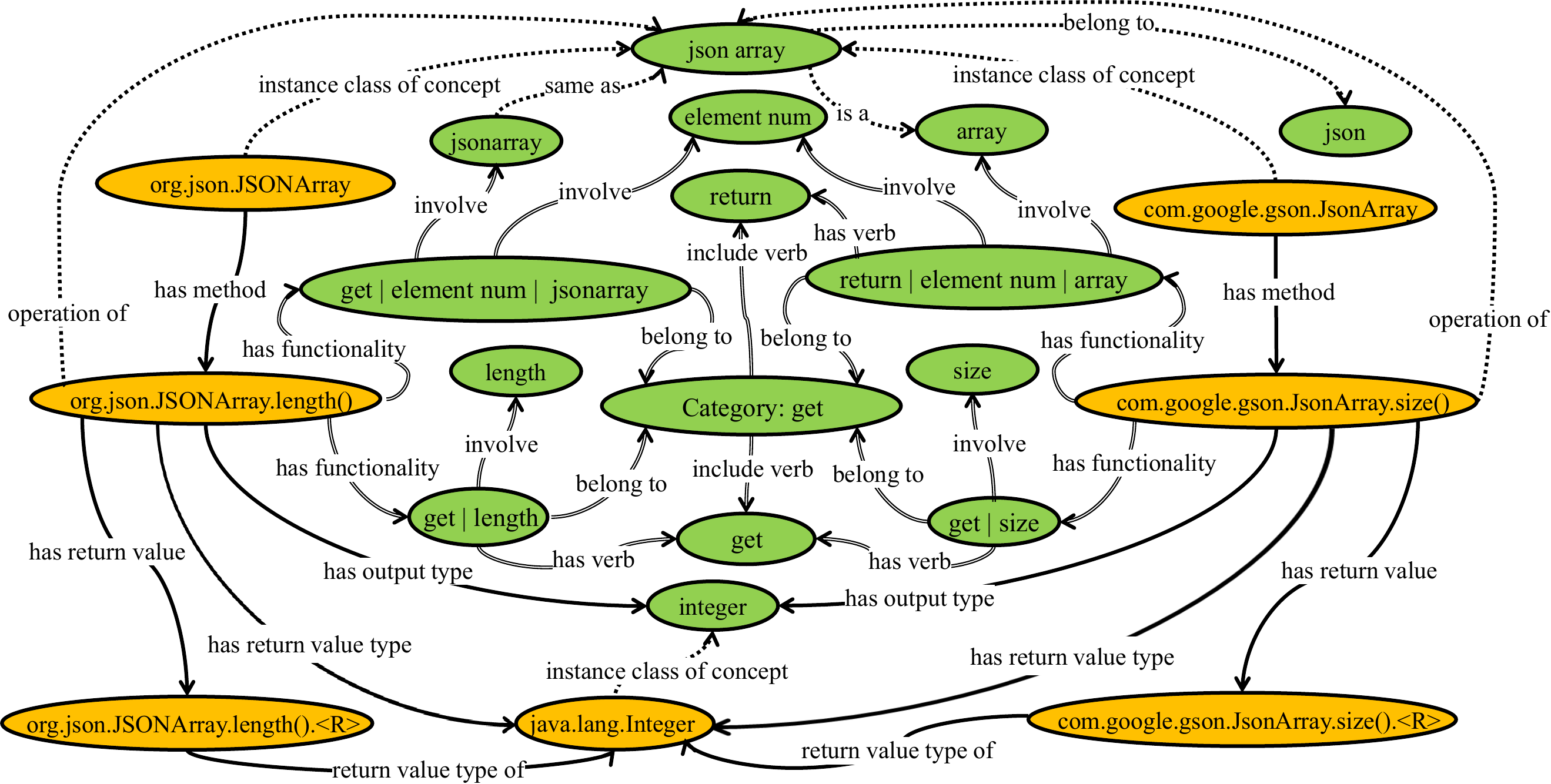}
	\vspace{-6mm}
	\caption{\textcolor{black}{An Example of API KG}}
	\vspace{-8mm}
	\label{fig:kgexample}
\end{figure}

\textcolor{black}{
Our API KG captures the structural and high-level information present in API documentation. It consists of entities (nodes) and relations (edges) that represent various aspects of APIs. Here, we offer definitions for key entities and relations:}

\begin{itemize}[itemsep=2pt,topsep=0pt,parsep=0pt, leftmargin=10pt]
    \item \textbf{API Element}. \textit{API elements} encompass components like libraries, packages, classes/interfaces, fields, methods, return values, parameters, and abstract parameters, forming the fundamental API building blocks.
    \item \textbf{Structural Relation}. \textit{Structural relations} describe the relationships between API elements, including ``extend'' (inheritance), ``implement'' (interface implementation), ``has field'' (fields within classes/interfaces), ``has method'' (methods within classes/inter\-faces), and ``has parameter'' (methods with required parameters), forming the API KG's foundation.

    \item \textcolor{black}{\textbf{Functionality Expression Element}. \textit{Functionality expression elements} pertain to the structural representation of API functionality descriptions. This includes functionality expressions, functionality verbs, functionality categories, phrase patterns. They facilitate the standardized representation of API functionalities, as defined by Xie et al.~\cite{fse20funcverb}.}

     \item \textcolor{black}{\textbf{Functionality Expression}.  A \textit{functionality expression} provides a structural representation for the functionality descriptions of methods following the standardized form defined by Xie \etal~\cite{fse20funcverb}. It is extracted from the description sentence of a method.}

    \item \textcolor{black}{\textbf{Functionality Verb}. A \textit{functionality verb} represents the verb that express the main action of the functionality, \eg ``return'', ``get'', and ``obtain''.}

    \item \textcolor{black}{\textbf{Functionality Category}. A \textit{functionality category} categorizes the functionality expressions based on their semantic meanings, which is abstracted from a set of functionality verbs that have similar meanings, \eg ``return'', ``get'', and ``obtain'' can be classified into the same category.}
    
     \item \textcolor{black}{\textbf{Phrase Pattern}. \textit{Phrase patterns} capture specific syntactic patterns or templates used in functionality expressions, \eg \textit{``V \{{patient}\}''} and \textit{``V \{{patient\}} in \{{location}\}''}.} \textcolor{black}{In the phrase pattern \textit{``V \{patient\} in \{location\}''} the placeholders ``{patient}'' and ``{location}'' represent noun phrases that fulfill semantic roles. ``\{Patient\}`` corresponds to the direct object of the functionality verb, signifying the entity or object directly affected by the action. ``\{location\}'' denotes the spatial or temporal context associated with the verb.}
     
      \item \textcolor{black}{\textbf{Concept}}. \textcolor{black}{Concepts in the API KG are specific semantic units that capture domain-specific knowledge or common themes in API documentation. These concepts are typically represented by noun phrases. For instance, in APIs related to JSON processing, concepts like JSON arrays, JSON objects, keys, and values frequently appear.
      Concepts may be involved in functionality expressions by playing some semantic roles (\eg patient, location).
      }
\end{itemize}

\textcolor{black}{
Figure~\ref{fig:kgschema} showcases the schema of our API KG, illustrating the types of entities and relations involved. Furthermore, Figure~\ref{fig:kgexample} provides a partial API KG example, highlighting the interconnectedness of these entities and relations. The complete schema, including definitions for all the entity and relation types, is available in our replication package~\cite{replication_package}.}
\textcolor{black}{
The orange ellipses and solid lines denote API elements and the structural relations between them, respectively.
Among them, abstract parameters represent the abstraction of the parameters of different methods that share the same names and types.
For example, all the method parameters with the name \textit{path} and type \textit{java.lang.String} are treated as the instances of the same abstract parameter.}
\textcolor{black}{
The green ellipses denote functionality expression elements (\ie functionality expressions, functionality verbs, functionality categories, and phrase patterns) and related concepts and double lines denote relations between these elements and concepts.}
Note that multiple methods may share the same functionality expression if their functionality descriptions include the same functionality verb, phrase pattern, functionality category, and concepts. 
\textcolor{black}{
The dashed lines denote various relations (\eg \textit{``is a''} and  \textit{``belong to''}) between concepts and the involvement relations between API elements and concepts (\eg \textit{``has input type''} between methods and concepts).
Some relations are omitted in Figure~\ref{fig:kgschema} for brevity, \eg the \textit{``instance class of concept''} relation between classes and concepts (see Section~\ref{sec:complete_relation}).}


In this way, API elements from different libraries can be indirectly related through structural relations (\eg shared parameter or return types), functionality expressions (\eg shared functionality categories and involved concepts), and concepts (\eg associated with related concepts).

Figure~\ref{fig:kgexample} shows some entities and relations related to the API methods \textit{org.json.JSONArray.length()} and \textit{com.google.gson.JsonArray\-.size()}.
The two methods are analogical API methods from two libraries \textit{org.json}~\cite{org.json} and \textit{gson}~\cite{gson}.
Although they have different names (\ie \textit{length()} and \textit{size()}) and functionality descriptions (\ie \textit{``Get the number of elements in the JSONArray, included nulls''} and \textit{``Returns the number of elements in the array''}), they are indirectly related in the API KG through different kinds of relations.
They share similarities such as return value type, functionality category, and associations with concepts like ``json array'' and ``element num''.

\subsubsection{Structure Knowledge Extraction}
\label{sec:structure_extraction}
This step extracts structure knowledge from the document so as to construct the basic skeleton of the API KG. In this work, we focus on Java libraries due to its popularity, but our approach is not specific to the programming language. We use the Javadoc API documentation in the JAR files of each library given its neat format and prevalence, and \app could also use API documentation from other sources (\eg online official documentation). In particular, \app extracts all API elements and their structural relations from the API definition according to the schema shown in Figure~\ref{fig:kgschema}. Meanwhile, \app{} further extracts the textual descriptions of API elements from their Javadoc comment (\ie the comment before the method declaration~\cite{doccomment}). The extracted text descriptions \textcolor{black}{can} be used for the subsequent functionality knowledge extraction and conceptual knowledge extraction.
\textcolor{black}{In our implementation, we utilize JavaParser~\cite{javaparser} to analyze the Java source files contained within JAR files. Through static analysis based on abstract syntax tree (AST), we extract all the API elements, as well as their structural relations and textual descriptions.
}

\subsubsection{Functionality Knowledge Extraction}
\label{sec:func_analysis}
We extract functionality knowledge of API methods by analyzing their names and text descriptions.
Xie \etal \cite{fse20funcverb} provide a dataset for standardized functionality description which is available online~\cite{funcverbnet-replication}.
It includes 10,016 functionality verbs, 89 functionality categories, and 523 phrase patterns.
We add all of them into the API KG as the basis of functionality knowledge extraction.
Xie \etal \cite{fse20funcverb} also provide a tool FuncVerbNet~\cite{funcverbnet}, which can parse a functionality description into a standardized functionality expression.
FuncVerbNet uses a text classifier to classify a functionality description into a functionality category and then identifies the corresponding phrase pattern, functionality verb, and concepts based on dependency tree parsing.
For example, it extracts the following functionality expression from the description \textit{``returns the number of elements in the array''}:

\noindent\fbox{
	\begin{minipage}{.95\linewidth}{
			\textbf{\textit{Functionality Category}}: get;
			
			\textbf{\textit{Functionality Verb}}: return;
			
			\textbf{\textit{Phrase Pattern}}: V \{patient\} in \{location\};
			
			\textbf{\textit{Concepts}}: [element number, array];
			
			\textbf{\textit{Functionality Expression}}: return | element number | array
		} 
\end{minipage}}
\vspace{1mm}



For each API method, we take the first sentence of its text description as its functionality description (if exists), following previous work~\cite{fse20funcverb,ase20apicomp}.
\textcolor{black}{
Next, we utilize FuncVebNet to extract the associated functionality expressions. The concepts present in the functionality expressions, which correspond to noun phrases that fulfill semantic roles in the phrase pattern, are extracted and refined through the removal of stop words and lemmatization techniques~\cite{fse20funcverb}. If the extracted functionality expressions and associated concepts do not already exist in the API KG, we add them as entities and establish \textit{``involve''} relations between them. We also establish relations between the extracted functionality expressions and other existing elements like functionality verbs, phrase patterns, and functionality categories defined by the schema (see Figure~\ref{fig:kgschema}).}

If a method has no text description,  we extract functionality expression from its name.
We split the name into a sequence of tokens according to camel case and underscore and then use the token sequence as the functionality description of the method.
For example, \eg \textit{``get Int''} can be extracted from the name of the method \textit{getInt()} as its functionality description.
If a verb is missing at the beginning of the method name, we add a default functionality verb according to the following rules. \textcolor{black}{We utilize WordNet~\cite{wordnet}, a lexical database that provides word meanings and classifications, to determine the part of speech (\eg adjective, noun) of words.}
\begin{itemize}[itemsep=2pt,topsep=0pt,parsep=0pt, leftmargin=10pt]
	\item Add \textit{``get''} if the method name is a noun phrase, \eg \textit{``get length''} for \textit{JSONArray.length()};
	\item Add \textit{``convert''} if the method name starts with ``to'', \eg \textit{``convert to String''} for \textit{JSONArray.toString()};
	\item Add \textit{``check''} if the method name is an adjective, \eg \textit{``check empty''} for \textit{ArrayList.empty()}.
\end{itemize}


\subsubsection{Conceptual Relation Completion}
\label{sec:complete_relation}
Conceptual relation completion establishes conceptual relations between analogical APIs by analyzing the names/descriptions of API elements and concepts and then completing conceptual relations for methods.
API element name/description analysis creates relations between API elements and concepts and adds new concepts if necessary.
Concept name analysis creates relations between concepts.
Method conceptual relation completion completes the relations between API methods and concepts based on existing relations. 

\parabf{API Element Name Analysis}.
Each API element (except method) can be regarded as an instance of a corresponding concept, for example \textit{java.io.File} represents an instance of the concept \textit{file}.
We extract the corresponding concepts in different ways according to the type of API elements:

\begin{itemize}[itemsep=2pt,topsep=0pt,parsep=0pt, leftmargin=10pt]
	\item Package, Class, and Interface: the lowercase phrase obtained by splitting the short name (\ie the part after the last dot of the fully qualified name) of the API element by camel case and underscore, \eg \textit{``json array''} is the concept for \textit{org.json.JSONArray};

	\item Return Value: the lowercase phrase obtained by splitting the return value type's short name by camel case and underscore;

	\item Parameter and Field: the lowercase phrase obtained by splitting the short name of the parameter/field by camel case and underscore \eg ``src file'' is the concept for \textit{File srcFile}.
\end{itemize}
For each concept obtained in this way we create an ``instance of'' relation between the API element and the concept, \eg \textit{<org.json.JSON\-Array, instance class of concept, json array>}.

\parabf{API Element Description Analysis}.
We extract concepts from the descriptions of API elements with the following steps:

\begin{itemize}[itemsep=2pt,topsep=0pt,parsep=0pt, leftmargin=10pt]
	\item Extract all the noun phases with Spacy~\cite{spacy}, for example \textit{``A JSONObject''} and \textit{``the value''} are extracted from the description of a return value \textit{``A JSONObject which is the value''};
	
	\item Lowercase and lemmatize extracted noun phrases, for example \textit{``files''} and \textit{``A JSONObject''} are converted into \textit{``file''} and \textit{``a jsonobject''}, respectively;

	\item Remove stop words at the beginning of a phrase, for example ``a'' is removed from \textit{``a jsonobject''}.
\end{itemize}

All the remaining noun phrases are treated as concepts mentioned in the description of API elements and the corresponding concept mention relations are created between them, \eg \textit{<jsonobject, mentioned in return value description, org.json.JSONObject.optJSON\-Object(java.lang.String).<R>}\textit{>}.

\textbf{Concept Name Analysis}.
The name of a concept may imply some conceptual relations between concepts, \eg \textit{<json array, is, array>}.
Such conceptual relations are useful for establishing possible associations between API elements with subtle differences in concept expression.
Following the previous work~\cite{ase20apicomp}, we use the following rules to identify possible conceptual relations between two concepts $C1$ and $C2$ in the API KG:

\begin{itemize}[itemsep=2pt,topsep=0pt,parsep=0pt, leftmargin=10pt]
	\item If $C1$'s name is derived from $C2$'s name, add a relation \textit{<C1, derived from, C2>}, \eg \textit{<builder, derived from, build>}.

	\item If $C1$'s name is shorter than and the prefix of $C2$'s name and there are no other longer concepts that satisfy this rule for $C1$, add
a relation \textit{<C2, facet of, C1>}, \eg \textit{<character sequence length, facet of, character sequence>};

	\item If $C1$'s name is shorter than the suffix of $C2$'s name and there are no other longer concepts that satisfy this rule for $C1$, add
a relation \textit{<C2, is, C1>}, \eg \textit{<json array, is, array>}.

	\item If $C1$'s name is the same as $C2$'s name after removing spaces, add bidirectional relations \textit{<C2, same as, C1>} and \textit{<C1, same as, C2>}, \eg \textit{<json array, same as, jsonarray>} and \textit{<jsonarray, same as, json array>};

\end{itemize}

\parabf{API Method Conceptual Relation Completion}. 
To better reflect the conceptual associations between methods in the subsequent API KG embedding, we further create direct relations between methods and concepts that are indirectly connected through multi-hop relations.
We follow the rules shown in Table~\ref{tab:r_supplement_rule} to complete the relations.
In this way, we establish direct relations between methods and concepts based on different parts of the methods, \ie object, input value, input type, and output type.

\begin{table}[]
	\centering
	\vspace{-2mm}
	\caption{Method Conceptual Relation Completion Rules (M: Method; C: Class, T: Type; Con: Concept)}
	\label{tab:r_supplement_rule}
	\footnotesize
	\vspace{-4mm}
	\begin{tabular}{|c|c|}
		\hline
		Existing Multi-hop Relations                                            & Completed Relation                                                  \\ \hline
		\textless{}C, has method, M\textgreater{}        & \multirow{2}{*}{\textless{}M, operation of, Con\textgreater{}}  \\
		 \textless{}C, instance class of concept, Con\textgreater{}                  &                                                                  \\ \hline
		\textless{}M, has parameter, P\textgreater{}                   & \multirow{2}{*}{\textless{}M, has input value, Con\textgreater{}}      \\
		\textless{}P, instance parameter of concept, Con\textgreater{} &                                                                  \\ \hline
		\textless{}M, has parameter type, T\textgreater{}              & \multirow{2}{*}{\textless{}M, has input type, Con\textgreater{}} \\ 
		\textless{}T, instance class of concept, Con\textgreater{}     &                                                                  \\ \hline
		\textless{}M, has return value type, T\textgreater{}           & \multirow{2}{*}{\textless{}M, has output type, Con\textgreater{}}     \\ 
		\textless{}T, instance class of concept, Con\textgreater{}     &                                                                  \\ \hline
	\end{tabular}
\vspace{-5mm}
\end{table}

\subsection{API Knowledge Graph Embedding}
\label{sec:kg_embed}

In this phase, \app trains a KG embedding model based on all the relation triples of the API KG. The model maps all the entities in the API KG (\eg API elements, functionality expression elements, concepts) to a high-dimensional vector space, where API elements with similar structural, functionality, and conceptual relationships are close. The benefits of KG embedding include: (i) graph embedding could well reserve both structural and semantic information in the graph, and (ii) mapping APIs into vector spaces could accelerate similar API retrieval since all API vectors are restored in a vector database and the vector index is very efficient.

In particular, we use the \textcolor{black}{ComplEx model}~\cite{complex}, a tensor decomposition based KG embedding method, to train the API KG embedding model.
A tensor decomposition models the  KG as a three-way tensor (\ie a three-dimensional adjacency matrix), which can be decomposed into a combination of low-dimensional vectors (\ie the embeddings of entities and relations~\cite{complex}).
ComplEx calculates a score for each relation triple \smalltexttt{<$h$,$r$,$t$>} using the equation: $\phi(h, r, t)=E_h \times E_r \times E_t$, where $h$, $r$ and $t$ are the head entity, relation type and tail entity respectively, and $E_h$, $E_r$ and $E_t$ are their  embeddings.
The score indicates the probability that the corresponding relation holds.
The model training takes all the relation triples in a KG as input and produces the embeddings of all the entities and relations in the KG as output.
\textcolor{black}{
The goal of the optimization during training is to assign a higher score to the true triplet $(E_h, E_r, E_t)$ compared to the corrupted false triplets $(E_h', E_r, E_t)$ and $(E_h, E_r, E_t')$.
}
To support antisymmetric relations, the model represents $E_h$, $E_r$ and $E_t$ in complex-valued space instance of real-valued space, \eg $h$ has a real part $Re(h)$ and an imaginary part $Im(h)$, \ie $h = Re(h)+iIm(h)$.

Given the large size of the API KG (\ie including more than 72 million entities and more than 289 million relations), we use PyTorch-BigGraph (PBG)~\cite{pbg} and its implementation shared on GitHub~\cite{BigGraphGithub} to train the ComplEx model. PyTorch-BigGraph is a distributed system implemented by Facebook with the purpose of supporting the training of knowledge graph embedding models on large graphs. We also investigate using other KG embedding models (\eg TransE~\cite{transE}  and DistMult~\cite{DistMult}) in Section~\ref{sec:rq3}.


To facilitate more efficient similarity calculation based on the KG embeddings, we store all the KG embeddings in a vector database, \ie Milvus~\cite{2021milvus}.
Milvus is an open-source vector database that supports high-efficient vector index and similarity search.
Based on Milvus, we can efficiently obtain the KG embeddings for a given entity in the KG or find the top-$k$ similar entity embeddings for a given embedding.

\begin{figure}
	\centering
	\vspace{-1mm}
	\includegraphics[width=0.77\columnwidth]{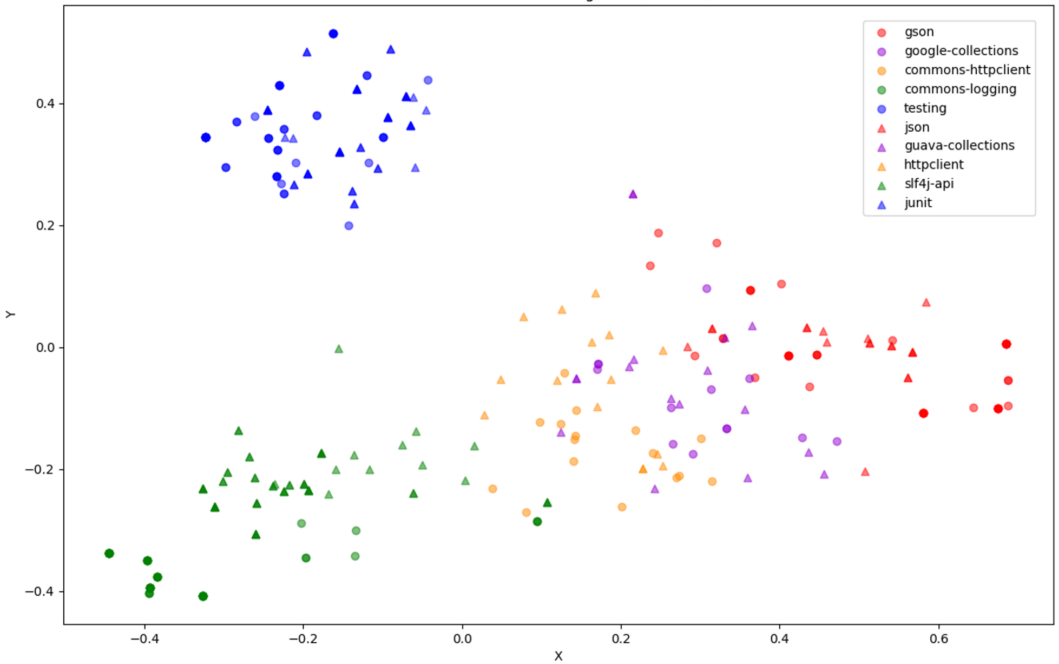}
	\vspace{-4mm}
	\caption{Examples of API KG Embeddings Using ComplEx}
	\vspace{-6mm}
	\label{fig:kg_emb_example}
\end{figure}

Figure~\ref{fig:kg_emb_example} shows the distribution of KG embeddings of some API methods in the vector space, which is generated after dimension reduction through PCA (Principal Component Analysis)~\cite{pca}.
Each point in Figure~\ref{fig:kg_emb_example} represents an API method from our benchmark (in  Section~\ref{sec:eva:Protocol}).
Points with the same color and shape (\ie triangle or circle) represent API methods from the same library.
The API methods of the two analogical libraries have the same color but different shapes.
We could observe that API methods in the same library (\eg \textit{org.json}) or analogical libraries (\eg \textit{org.json} and \textit{gson}) are relatively close in the vector space, while the API methods of libraries with different topics are far apart.
For example, the API methods of the libraries related to  logging (\textit{slf4j}~\cite{slf4j} and \textit{commons-logging}~\cite{commons-logging}) are far apart from the ones of the libraries related to testing (\eg \textit{junit}~\cite{junit} and \textit{org.testing}~\cite{org.testing}).

\subsection{Analogical API Method Inferring}
\label{sec:app_infer}
In this phase, \app returns a list of ranked analogical API methods for a given source API method based on the API KG and the embedding model.
First, \app selects candidate API methods based on their similarities with the given API method (candidate API method retrieval in Section~\ref{sec:vector_retrieval}); then, \app re-ranks the candidate API methods by considering the similarities between the given API method and the neighbors of the candidate API methods (candidate API method re-ranking in Section~\ref{sec:candidate_rerank}). The purpose of the candidate API method retrieval in the first step is to narrow down the scope of candidate APIs, so that the second re-ranking step only needs to calculate the similarity between the given API and a small number of candidate APIs. 



\subsubsection{Candidate API Method Retrieval}
\label{sec:vector_retrieval}
For a given source API method $s$, we first obtain its KG embedding $E_s$ by \textcolor{black}{querying Milvus}.
\textcolor{black}{Then we calculate the KG similarity $Sim_{kg}$ between $s$ and methods from other libraries (called method similarity $Sim_{m}$) according to Eq.~\ref{eq:sim:kg}, which is normalized cosine similarities between their KG embeddings.}
\textcolor{black}{We select the top-$k$ (\eg 100) API methods as candidates, utilizing the efficient vector indexing in our database, which achieves low latency in milliseconds on trillion vector datasets.}
\begin{equation}
	\vspace{-2mm}
	\label{eq:sim:kg}
	\footnotesize
	Sim_{kg}(E_1,E_2)=(Cos(E_1,E_2)+1)/2
\end{equation}

\subsubsection{Candidate API Method Re-ranking}
\label{sec:candidate_rerank}
Two API methods with high KG embedding similarity are not necessarily analogical API methods. 
For example, \textit{org.json.JSONArray.getJSONObject(int)} and \textit{com.google.gson.JsonArray.remove(int)} have a high KG embedding similarity since they belong to analogical classes.
To address this issue, we further compute the similarity between the same type of neighbor concepts of the given API method $s$ and each candidate API method $e$, which reflects the conceptual similarity of API methods in different aspects (\eg functionalities, inputs, and outputs).
The neighbor-based similarities we compute include functionality similarity $Sim_{func}$, object similarity $Sim_{obj}$, input type similarity $Sim_{it}$, input value similarity $Sim_{iv}$, output type similarity $Sim_{ot}$, and average neighbor similarity $Sim_{neig}$.
To get the final analogical score $Score(s,e)$, we then perform a weighted sum of these neighbor-based similarities and the method similarity $Sim_{m}$ according to Eq.~\ref{eq:sim:sum}.
\begin{equation}
	\vspace{-1mm}
	\label{eq:sim:sum}
	\footnotesize
	Score(s,e)= \sum_{t\in {m,func,obj,iv,it,ot,neig}} W_{t} \times Sim_{t}(s,e)
\end{equation}

\textcolor{black}{
All candidates are ranked by the analogical scores. We then explain each similarity as follows.}

\textbf{Method Similarity ($Sim_{m}$}).
 $Sim_{m}$ is the KG similarity between two methods, which is already computed in the retrieval step.

\textbf{Functionality Similarity ($Sim_{func}$)}.\textcolor{black}{
The functionality similarity measure ($Sim_{func}$) captures the similarity in the functionalities provided by two API methods. It relies on the assumption that comparable APIs should have similar functionality expressions.}
We calculate the maximum similarity between the functionality expressions corresponding to the two methods according to Eq.~\ref{eq:sim:func} as their functionality similarity $Sim_{func}(s,e)$.
In Eq.~\ref{eq:sim:func}, $Func(s)$ denotes the functionality expression of the method $s$ (\ie \textit{<$s$, has functionality, $Func(s)$>}), which is extracted from the method name or the functionality description (see Section~\ref{sec:func_analysis}).
\textcolor{black}{This measure allows us to capture the similarity of API methods based on their intended functionality and purpose.}
\begin{equation}
	\vspace{-1mm}
	\label{eq:sim:func}
	\footnotesize
	Sim_{func}(s,e)=Max(Sim_{kg}(E_{Func(s)},E_{Func(e)}))
\end{equation}

\textbf{Object Similarity ($Sim_{obj}$)}.
\textcolor{black}{$Sim_{obj}$ captures the conceptual-level similarity between the classes of two API methods. It is based on the intuition that methods belonging to analogous classes are likely to exhibit similar behavior and usage patterns.}
$Sim_{obj}$ is calculated according to Eq.~\ref{eq:sim:object}, where $Obj(s)$ represents the concept corresponding to the class of the method $s$ (\ie \textit{<$Obj(s)$, has operation, $s$>}).
\begin{equation}
	\vspace{-1mm}
	\label{eq:sim:object}
	\footnotesize
	Sim_{obj}(s,e)=Sim_{kg}(E_{Obj(s)},E_{Obj(e)}))
\end{equation}

\textbf{Input Type Similarity ($Sim_{it}$)}.
 $Sim_{it}$ of two methods reflects the conceptual-level similarity of their parameter types.
 \textcolor{black}{Analogical APIs are expected to operate on similar types of input data.}
$Sim_{it}$ is calculated according to Eq.~\ref{eq:sim:it}, where $InType(s)$ represents a concept corresponding to one of the parameter types of the method $s$ (\ie \textit{<$InType(s)$, has input type, $s$>}) and $\bar E_{InType(s)}$ represents the average of KG embeddings of all $InType(s)$.
\begin{equation}
	\vspace{-1mm}
	\label{eq:sim:it}
	\footnotesize
	Sim_{it}(s,e)=Sim_{kg}(\bar E_{InType(s)},\bar E_{InType(e)}))
\end{equation}

\textbf{Input Value Similarity ($Sim_{iv}$).}
 \textcolor{black}{The purpose of $Sim_{iv}$ is to capture the conceptual-level similarity of parameters between two methods, which contributes to identifying analogical APIs. Analogical APIs often exhibit similarities in the values they accept as input, irrespective of the specific parameter types.}
$Sim_{iv}$ is calculated according to Eq.~\ref{eq:sim:iv}, where $InVal(s)$ represents a concept corresponding to one of the parameter of the method $s$ (\ie \textit{<$InVal(s)$, has input value, $s$>}) and $\bar E_{InVal(s)}$ represents the average of KG embeddings of all $InVal(s)$.
\begin{equation}
	\vspace{-1mm}
	\label{eq:sim:iv}
	\footnotesize
	Sim_{iv}(s,e)=Sim_{kg}(\bar E_{InVal(s)},\bar E_{InVal(e)}))
\end{equation}

\textbf{Output Type Similarity ($Sim_{ot}$)}.
 $Sim_{ot}$ of two methods reflects the conceptual-level similarity of their return value types.
 \textcolor{black}{Analogical APIs often exhibit similarities in the types of values they return.}
$Sim_{ot}$ is calculated according to Eq.~\ref{eq:sim:object}, where $OutType(s)$ represents a concept corresponding to the return value type of the method $s$  $s$ (\ie \textit{<$s$, has output type, $OutType(s)$>}).
\begin{equation}
	\vspace{-1mm}
	\label{eq:sim:ot}
	\footnotesize
	Sim_{ot}(s,e)=Sim_{kg}(E_{OutType(s)},E_{OutType(e)})
\end{equation}

\textbf{Average Neighbor Similarity ($Sim_{neig}$)}.
\textcolor{black}{Analogical APIs often exhibit similarities not only in their individual aspects but also in their overall context or behavior. By calculating $Sim_{neig}$ using Eq.\ref{eq:emb:overall} and Eq.\ref{eq:sim:overall_beh}, where $E_{Neig(s)}$ represents the average of KG embeddings of the method and its neighboring concepts, we can capture the similarity of overall neighbors between two methods. This similarity measure provides a holistic view of the methods' surrounding context, allowing us to identify analogical APIs based on the similarity of their overall behavior.}
\begin{equation}
		\vspace{-1mm}
		\label{eq:emb:overall}
		\footnotesize
		\begin{split}
			E_{Neig(s)}=Avg(E_{s}+E_{Obj(s)}+E_{Func(s)}\\+\bar E_{InVal(s)}+\bar E_{InType(s)}+E_{OutType(s)})
		\end{split}
\end{equation}
\begin{equation}
		\vspace{-3mm}
		\label{eq:sim:overall_beh}
		\footnotesize
		Sim_{neig}(s,e)=Sim_{kg}(E_{Neig(s)}, E_{Neig(e)}))
		\vspace{3mm}
\end{equation}

Note that instead of directly using the similarity of neighboring API elements of two methods (\eg their return values), we use concepts related to neighboring API elements, as API elements are library-specific while concepts are more likely to be shared between libraries. In addition, to ensure the diversity of the return APIs, we further limit the number (\ie 3) of recommended API methods that come from the same library.




\section{Evaluation}
\label{sec:evaluation}

\begin{table}[]
	\footnotesize
	\centering
 	\vspace{-2mm}
	\caption{Statistics of Resulting API KG}
	\vspace{-4mm}
	\label{table:kg_statistics}
	\begin{tabular}{|l|r|l|r|}
		\hline
		Type          & Number & Type          & Number \\ \hline
		Library &   35,773 & Return Value          & 15,451,223   \\ \hline
		Package &   229,061 & Abstract Parameter & 1,892,120  \\ \hline
		Class &   3,090,537 & Functionality Expression &   5,200,297  \\ \hline
		Interface &    281,854 &   Functionality Category &  89    \\ \hline
		Field &   6,232,643   & Functionality Verb  &  10,016  \\ \hline
		Method &    15,441,057  & Phrase Pattern &  523\\ \hline
		Parameter          & 16,501,363&  Concept &    5,660,553 \\ \hline
	\end{tabular}
\vspace{-6mm}
\end{table}
To implement \app, we construct a unified API KG from 35,773 Java libraries. 
\textcolor{black}{Table~\ref{table:kg_statistics} presents the entity type statistics of the resulting API KG. To collect the Javadoc documentation for those libaries,} we first get the metadata (\eg groupId and artifactId) of a list of Java libraries according to the Libraries.io dataset \cite{libariesiodata} (last updated in January 2020); then, we download the latest version of JAR files (as of August 11, 2022) from the Maven Central Repository, resulting in 35,773 JAR files; lastly, we leverage zipfile~\cite{zipfile} and JavaParser~\cite{javaparser} to extract the API-relevant documentation from JAR files, including the API definition and API functionality descriptions.
In this way, we construct an API KG with 72,242,099 entities, including 59,155,631 API elements, 5,210,925 functionality elements, and 5,660,553 concepts.
Further, we train the KG embedding model using ComplEx with a logistic loss.

\textcolor{black}{
The weight of each similarity in Eq.~\ref{eq:sim:sum} is determined as  $W_{m}=0.05$, $W_{func}=0.95$, $W_{obj}=0.8$, $W_{it}=0.25$, $W_{iv}=0.05$, $W_{ot}=0.05$, and $W_{neig}=0.95$ based on our experiments in a separate validation setting (to avoid overfitting) and we also investigate the impact of different weights in Section~\ref{sec:rq3}.}

We evaluate \app by answering the following research questions. 
RQ1 and RQ2 investigate the effectiveness of \app in two analogical API recommendation scenarios, \ie one with the given target library and the other without the given target library. 
To better understand the capabilities and characteristics of \app, RQ3 analyzes the impact of different components in \app, and RQ4 further studies the scalability of \app when the number of libraries is increasing.


\begin{itemize}[itemsep=2pt,topsep=0pt,parsep=0pt, leftmargin=10pt]
\item \textbf{RQ1 (Effectiveness with target libraries)}: 
How does \app compare to existing documentation-based techniques when recommending analogical API methods \textit{with given target libraries}?
\item \textbf{RQ2 (Effectiveness without target libraries)}: 
How does KGE4\-AR compare to existing documentation-based techniques when recommending analogical API methods \textit{without given target libraries}?
\item  \textbf{RQ3 (Impact Analysis)}: How do different components in KGE4\-AR (i.e., the KG embedding models, knowledge types, and similarity types and weights) impact the effectiveness of \app? 
\item \textbf{RQ4 (Scalability)}: How scalable is KGE4AR with the increasing number of libraries?
\end{itemize}

\subsection{RQ1: Effectiveness with Target Libraries}
\label{sec:rq1}

In this RQ, we evaluate the effectiveness of \app and state-of-the-art documentation-based analogical API recommendation techniques with given target libraries.
\textcolor{black}{
}

\subsubsection{Protocol}\label{sec:eva:Protocol}
\textcolor{black}{In this section, we introduce the benchmark, baselines, and metrics utilized for this research question.}

\parabf{Benchmark.} 
There are two exiting benchmarks~\cite{wcre13funcmap,icpc19apimap} of manually validated analogical API pairs; and we directly obtain both datasets from their replication packages~\cite{Teytondataset,Alrubayedataset} and merge them into one benchmark. In this way, we construct a large benchmark, which contains 245 pairs of analogical API methods from 16 pairs of analogical libraries, covering different topics such as JSON processing, testing, logging, and network requests. 
\textcolor{black}{For each analogical API pair, either API can be used as the source API, resulting in 490 source APIs (245 pairs $\times$ 2). In each query, the source API and all candidate APIs from the target library are provided as inputs, and the output is the ranked list of candidate APIs.}

\parabf{Baselines.} 
We include two state-of-the-art documentation-based analogical API recommendation techniques (\ie RAPIM~\cite{asc20apimigration} and D2APIMap~\cite{icpc20apimigration}) for comparison.  
We select these two techniques since they are the latest and the most effective ones in the unsupervised learning-based category and supervised learning-based category, respectively.

\begin{itemize}[itemsep=2pt,topsep=0pt,parsep=0pt, leftmargin=10pt]
    \item RAPIM~\cite{asc20apimigration} is a supervised learning-based approach, which trains a machine learning model (\ie boosted decision tree) and leverages the trained model to predict the probability of an unseen API pair being analogical. In particular, for a given API pair, RAPIM calculates a set of features that are based on the lexical similarity of the method descriptions, return type descriptions, method names, and class names between two APIs. We collect their features according to the paper and then directly use RAPIM via its network requests~\cite{asc_service}.


    \item D2APIMap~\cite{icpc20apimigration} is an unsupervised learning-based approach that utilizes the Word2Vec model to compute similarities between functionality descriptions, return values, and parameters of API pairs. It recommends the API with the highest total similarity. Due to the unavailability of the source code, we re-implement D2APIMap following the original paper.


\end{itemize}


\parabf{Metrics.} 
\textcolor{black}{Following prior work~\cite{tse19apimigration}, we adopt common evaluation metrics: MRR (Mean Reciprocal Rank) and Hit@k ($k=1,3,5,10$). MRR calculates the average rank of the correct analogical API in the generated list, while Hit@k measures the proportion of queries in which the correct analogical API appears within the top-k positions. Considering the vast number of APIs in each library, we limit our analysis to the top 100 candidates in the ranked list for each query.}


\begin{table}[]
	\centering
	\vspace{-2mm}
	\caption{Effectiveness with Given Target Libraries}
	\label{tab:rq1:effiveness:single_lib}
	\footnotesize
	\vspace{-4mm}
		\begin{tabular}{|c|c|c|c|p{0.8cm}<{\centering}|p{0.8cm}<{\centering}|}
		\hline
		Approach     & MRR & Hit@1 & Hit@3 & Hit@5 & Hit@10  \\ \hline
		RAPIM     &   0.158 &   0.082 &   0.180 &   0.229  &  0.304  \\ \hline
		D2APIMap  & 0.261   &  0.180   & 0.278  &  0.343  &  0.420  \\ \hline
		\app  & \textbf{0.384} &   \textbf{0.267} &   \textbf{0.449} &   \textbf{0.527}&    \textbf{0.594} \\ \hline
	\end{tabular}
	\vspace{-5mm}
\end{table}

\subsubsection{Results}
Table~\ref{tab:rq1:effiveness:single_lib} presents the evaluation results, and the best value of each metric is in boldface. 
\app substantially outperforms both baselines on all metrics. In particular, \app achieves 47.1\%-143.0\%, 48.3\%-225.6\%, 61.5\%-149.4\%, 53.6\%-130.1\%, and 41.4\%-95.4\% improvements over the baselines in terms of MRR, Hit@1, Hit@3, Hit@5, and Hit@10, respectively.

We further investigate the results and find the potential reason why \app outperforms baselines might be that \app analyzes API functionality descriptions in a better way.  
For example, when two APIs share the same noun phrases and different verbs (\eg \textit{StorageObject.getContentLength()} and \textit{S3ObjectWrapper.setObject\-Content(S3ObjectInputStream)}), it is often difficult for RAPIM and D2APIMap to distinguish them. RAPIM incorporates a TF-IDF model to calculate similarity-related features, which often assigns functionality verbs with low weights due to their high frequency in the names and descriptions; 
D2APIMap incorporates a Word2Vec model to calculate similarities, which often represents functionality verbs with similar vectors due to their similar contexts. 
However, \app extracts the functionality knowledge of methods (\eg functionality category, functionality verb), and considers functionality similarity of methods in the re-ranking step (see Section~\ref{sec:app_infer}), which can effectively distinguish the difference between methods even if they share same noun phrases. Therefore, in this example, \app successfully identifies these two APIs as not analogical while baselines consider them as analogical. 

In summary, \app substantially outperforms state-of-the-art documentation-base techniques when inferring analogical API methods with given target libraries. 
 


\subsection{RQ2: Effectiveness without Target Libraries}
\label{sec:rq2}
\textcolor{black}{RQ1 evaluates analogical API recommendation techniques when the target library is known. However, in practice, selecting the correct target library is challenging, and existing automated target library recommendation approaches have limited effectiveness (Top1-recall < 20\%~\cite{saner21libmigration}). Therefore, in this RQ, we assess the effectiveness of \app{} in the scenario where no target library is available.}


\subsubsection{Protocol}
We then introduce the benchmark, metrics, and baselines used in this RQ.

\parabf{Benchmark.}  The benchmark in RQ1 only contains analogical API pairs whose candidate APIs are from one given target library, and is not suitable for the analogical API recommendation scenario without target libraries. Therefore, in this RQ, we manually construct a new benchmark of analogical API pairs whose candidate APIs are from a wide range of libraries instead of from one given target library. \textcolor{black}{In particular, based on previous work~\cite{tse19apimigration}, online resources such as Awesome-Java~\cite{awesomejava}, and our expert knowledge, we first manually select 9 pairs of analogical libraries (\ie 18 libraries); then for each of these 18 libraries, we randomly select 15 API methods in the library as the source APIs for evaluation, leading to 270 source APIs in total. 
The selected libraries include both popular ones (usage number > 500 in Maven Central~\cite{mavencenter}), such as \textit{gson}~\cite{gson}, and less popular libraries, such as \textit{dsl-json}~\cite{dsl-json} and \textit{dom4j}~\cite{dom4j}.
The selected libraries represent diverse domains such as data processing and code analysis, ensuring the evaluation of our approach's effectiveness and generalizability in real-world scenarios.}

\textit{Ground-truth labeling.} 
We manually label whether the API pair in our newly-constructed benchmark is analogical or not. \textcolor{black}{Due to the large number of potential API pairs, we only label the Top-10 APIs returned by each technique in each query, resulting in a total of 6,986 labeled API pairs.} In particular, six participants each with more than three years of Java development experience manually assess whether the returned APIs are analogical to the source API.  In each query, two participants are asked to read the API documentation of the source API and the returned APIs to make the judgment whether they are analogical or not.
\textcolor{black}{The returned APIs for each source API are shuffled before assessment, and annotators are unaware of the technique that produced the results. In cases where the assessment by two annotators is inconsistent, a third annotator is involved to make a judgment, and the final annotation is based on majority agreement. The inter-annotator agreement is substantial, with a Cohen's Kappa coefficient~\cite{kappa} of 0.666.}

\parabf{Metrics.}
In addition to the four metrics used in RQ1 (i.e., MRR, Hit@1, Hit@3, Hit@5, and Hit@10), we further include precision and recall in this RQ, since in this scenario there could be multiple correct answers corresponding to a source API. 
In particular,  precision is the fraction of analogical API methods among the returned results, while recall is the fraction of analogical API methods that are retrieved. In total, we compare \app with baselines on all these metrics based on manually labeled ground truths.

\parabf{Baselines.}
Existing baselines (\ie RAPIM and D2APIMap) exhaustively calculate the similarity between the source API and all candidate APIs, and thus it is unaffordably expensive to directly apply these techniques when there is no given target library and the number of candidate APIs is extremely large (\eg there could be over 15 million candidate APIs for each source API when there is no specified target library). Therefore, in this RQ, we enhance baselines by first narrowing the scope of their candidate APIs. In particular, we first leverage the lightweight information retrieval technique BM25~\cite{bm25} to select Top-100 candidate APIs whose documentations share high relevance to the source API; we then apply baselines on these candidate APIs.
\textcolor{black}{We adopt BM25 for its effectiveness and efficiency~\cite{bm25}. Additionally, we clean the documentation (\eg removing stop words, splitting camel case, and performing lemmatization) following previous work~\cite{icpc20apimigration} to further enhance the effectiveness of BM25.}
For distinction, we denote two baselines (\ie RAPIM and D2APIMap) enhanced with BM25 as \newRAPIM{} and \newDAPIMap{}, respectively. 
We implement the BM25-based candidate selection with Elasticsearch~\cite{elasticsearch}.

\subsubsection{Results}
\begin{table}[]
 	\vspace{-2mm}
	\centering
	\caption{Effectiveness without Given Target Libraries}
	\label{tab:rq1:effiveness:cross_lib}
	\footnotesize
	\vspace{-3mm}
	\begin{tabular}{|p{1.05cm}<{\centering}|p{0.45cm}<{\centering}|p{0.6cm}<{\centering}|p{0.6cm}<{\centering}|p{0.6cm}<{\centering}|c|p{0.8cm}<{\centering}|p{0.55cm}<{\centering}|}
		\hline
		Approach    & MRR & Hit@1 & Hit@3 & Hit@5 & Hit@10 &Precision & Recall  \\ \hline
		\newRAPIM{}  & 0.381  & 0.311&0.404&0.485&0.585& 0.271   &   0.237       \\ \hline
		\newDAPIMap{} & 0.616    & 0.570& 0.644 &0.685&0.715& 0.369   &   0.385         \\ \hline
		\app   &  \textbf{0.688}  &  \textbf{0.648} & \textbf{0.719} & \textbf{0.737} & \textbf{0.774} & \textbf{0.513}  & \textbf{0.480}     \\ \hline
	\end{tabular}
	\vspace{-1mm}
\end{table}


Table~\ref{tab:rq1:effiveness:cross_lib} presents the evaluation results. Overall, \app outperforms both baselines on all metrics by achieving 11.7\%-80.6\%, 13.7\%-108.3\%, 11.6\%-77.9\%, 7.6\%-52.0\%, 8.3\%-32.3\%, 26.2\%-72.0\%, and 33.2\%-116.5\% improvements in terms of MRR, Hit@1, Hit@3, Hit@5, Hit@10, precision, and recall, respectively.

\begin{figure}
	\centering
	\vspace{-1mm}
	\includegraphics[width=0.66\columnwidth]{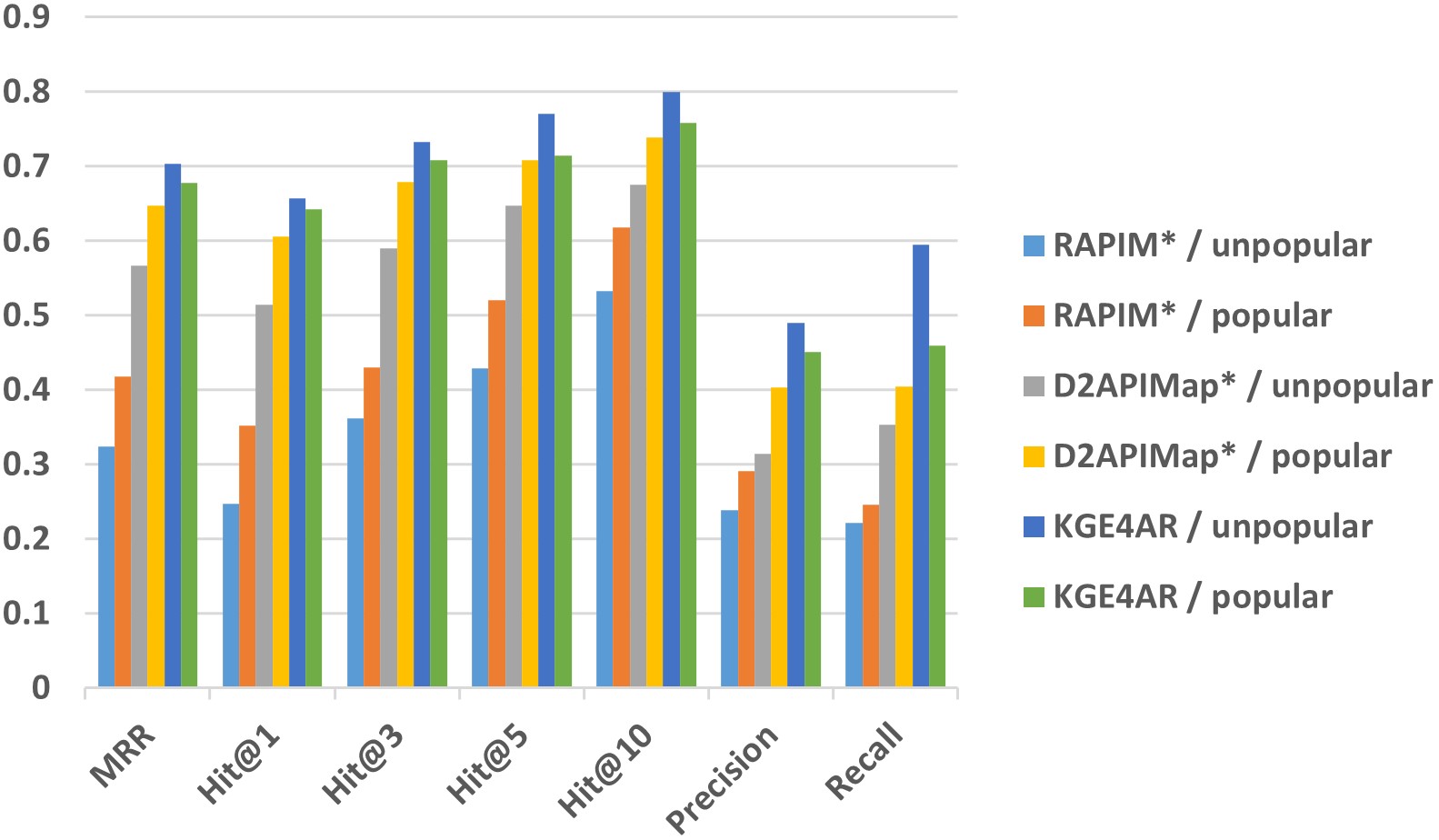}
	\vspace{-5mm}
	\caption{Effectiveness on Popular and Less Popular Libraries}
	\label{fig:pop-com}
\end{figure}

We further investigate how \app performs on different libraries. Figure~\ref{fig:pop-com} shows how \app and baselines perform on popular libraries and less popular libraries. We could find that \app consistently outperforms baselines in both popular and less popular libraries. Interestingly, the improvement of \app over baselines is even larger on those less popular libraries. For example,  MRR, precision, and recall of \app on \textit{dsl-json}~\cite{dsl-json} (with only 18 usages on Maven Central) are 0.542, 0.327, 0.562, respectively; while these metrics of \newDAPIMap{} on the same library are only 0.206, 0.080, and 0.171, respectively. 
One potential reason might be that the APIs of less popular libraries may target relatively uncommon functionality, whose descriptions may have a large semantic gap with analogical APIs. 
Existing baselines rely on simplistic text matching to recommend analogical APIs, which cannot handle less popular APIs well; while \app could better combine the structural information and functionality descriptions of APIs together through knowledge graph embedding to infer analogical APIs from a large number of candidate APIs.

In summary, \app outperforms existing techniques for inferring analogical APIs without given target libraries.

\subsection{RQ3: Factor Impact}
\label{sec:rq3}
In this RQ, we further analyze the impact of components in \app, including the re-ranking component, KG embedding models, knowledge types,  similarity types, and weights.  
Given the large number of comparison experiments in this RQ (\ie 15 runs), we perform experiments on a small-scale API KG based on RQ1 benchmark.

\subsubsection{Impact of Re-ranking Component} \label{sec:rerankimpact}
To investigate the contribution of the re-ranking step in \app{}, we include a variant (denoted as \app-Ret) of \app by removing the re-ranking step in inferring the analogical API.
The results of \app-Ret in MRR, Hit@1, Hit@3, Hit@5, and Hit@10 are 0.233, 0.133, 0.253, 0.327, and 0.447 respectively, which are much lower than the default \app (\eg 50.2\% lower in Hit@1). Such results indicate the re-rank step indeed contributes to the effectiveness of \app. 

\subsubsection{Impact of KG Embedding Models}
We train various KG embedding models on the small-scale API KG to explore their impact. We compare ComplEx with TransE~\cite{transE} and DistMult~\cite{DistMult}.
We evaluate KG embedding models using \app-Ret baseline on inferring analogical API methods with given target libraries (Section~\ref{sec:rq1}).
\app-Ret retrieves analogical API methods using KG embedding similarity, reflecting how well models learn method semantics.
Comparison is on top 100 results (Table~\ref{tab:kge}).
As shown in the table, ComplEx, the default in \app, achieves the best performance on all metrics, implying its suitability.

\begin{table}
	\centering
 	\vspace{-2mm}
	\caption{Impact of Different KG Embedding Models}
	\label{tab:kge}
	\footnotesize
	\vspace{-3mm}
	\textbf{}	\begin{tabular}{|p{2cm}<{\centering}|c|c|c|p{0.7cm}<{\centering}|p{0.8cm}<{\centering}|}
		\hline
		Method     & MRR & Hit@1 & Hit@3 & Hit@5 & Hit@10\\ \hline
		TransE    &  0.284  &  0.174    &  0.312    &  0.396   &   0.518     \\ \hline
		DistMult  &  0.288   &     0.180  &  \textbf{0.331}   & 0.400     &  0.494      \\ \hline
		ComplEx   &  \textbf{0.293}   &   \textbf{0.183}  & 0.324    &\textbf{0.422}   &   \textbf{0.524}   \\ \hline
		
	\end{tabular}
\vspace{-3mm}
\end{table}

\subsubsection{Impact of Knowledge Types in the API Knowledge Graph}
To evaluate the impact of different types of knowledge in the API KG, we train different KG embedding models based on a subset of relation triples in the small-scale API KG.
We try three situations: only structural relation triples (denoted as \textit{Structure}), all relation triples except functionality-related relations (denoted as \textit{Functionality*}), and all relation triples except conceptual relations (denoted as \textit{Concept*}).
Then we evaluate different KG embedding models based on \app-Ret and the benchmark as well.
The results are shown in Table~\ref{tab:knowledge_type}.
Both functionality and conceptual contribute positively to analogical API method inferring, while conceptual knowledge has a greater impact than functionality knowledge.


\begin{table}[]
	\centering
	\vspace{-0mm}
	\caption{Impact of Different Knowledge Types}
	\label{tab:knowledge_type}
	\footnotesize
	\vspace{-4mm}
	\begin{tabular}{|c|c|c|c|p{0.6cm}<{\centering}|p{0.8cm}<{\centering}|}
		\hline
		Knowledge Type     & MRR & Hit@1 & Hit@3 & Hit@5 & Hit@10  \\ \hline
		Structure         &  0.154   &  0.070     &   0.169    &   0.233    &     0.331          \\ \hline
		Functionality*  & 0.282    &  \textbf{0.185}   &   0.309     &  0.385     &       0.489        \\ \hline
		Concept*  &   0.237  &    0.150   &   0.259    &   0.335    &      0.422      \\ \hline
		All   &  \textbf{0.293}   &   0.183  & \textbf{0.324}    &\textbf{0.422}   &   \textbf{0.524}   \\ \hline
	\end{tabular}
\end{table}

\subsubsection{Impact of Similarity Types and Similarity Weights}
\begin{figure}
	\centering
	\vspace{-5mm}
	\includegraphics[width=0.77\columnwidth]{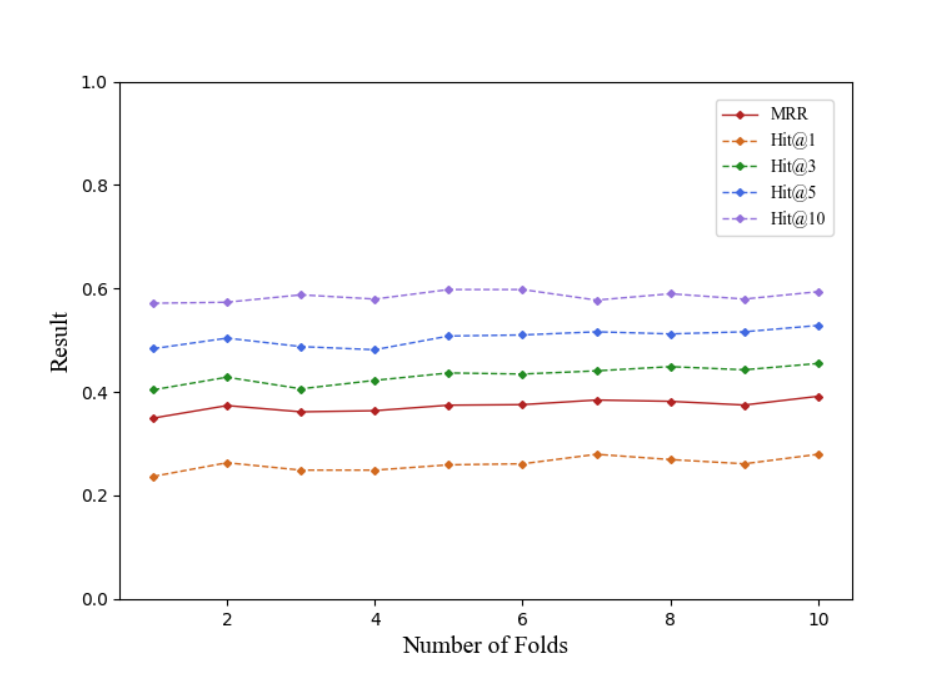}
	\vspace{-5mm}
	\caption{\textcolor{black}{Impact of the Number of Data for Tuning Weights}}
	\label{fig:fold_num}
\end{figure}

\begin{figure}
	\centering
	\vspace{-6mm}
	\includegraphics[width=0.58\columnwidth]{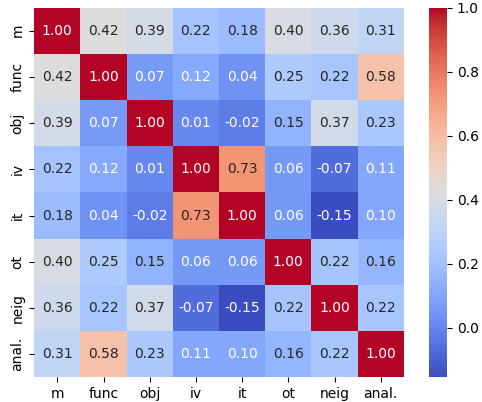}
	\vspace{-5mm}
	\caption{Similarities and Analogical Relationships Correlation Matrix}
	\label{fig:heapmap}
    \vspace{-5mm}
\end{figure}

As mentioned in Section~\ref{sec:candidate_rerank}, we tune the weights of similarities (\ie $W_{m}$, $W_{func}$, $W_{obj}$, $W_{it}$, $W_{iv}$, $W_{ot}$, and $W_{neig}$) in the re-ranking step on a small-scale API KG instead of on the large-scale API KG to avoid overfitting. In particular, we randomly divide the benchmark into 10 folds and then use a different number of folds to tune the weights in turn. We use Beam search~\cite{beamsearch} to tune all weights one by one with a step size of 0.05 and a beam number of 4.
Figure~\ref{fig:fold_num} shows experimental results of weights tuned with different folds of data. We could observe there is a subtle improvement when more tuning data is used, indicating that tuning with a small set of data might already be sufficient to achieve decent effectiveness. Note that our weight tuning is performed on a small-scale API KG, while the previous experiments (RQ1 and RQ2) are based on a large-scale API KG. Thus, it further indicates the tuned weights can be generalized even on different API KGs.
In addition, we further remove each similarity (by setting its weight as 0) so as to investigate its impact on the effectiveness of \app. 
Table~\ref{tab:rerank_sim} presents the evaluation results, with $Sim_{t}*$ representing the \app variant that excludes the similarity $Sim_{t}$. We can observe a decrease in the performance of \app when each similarity is removed. Particularly, the removal of functionality similarity $Sim_{func}$ leads to the largest drop, with a 22.9\% decrease in MRR. 
It shows the importance of the functionality knowledge for analogical API method inferring.
Additionally, removing $Sim_{neig}$ increases MRR and decreases Hit@10, suggesting that neighbor similarity brings some noise but improves recall.


\textcolor{black}{Figure~\ref{fig:heapmap} presents a heatmap of the correlation matrix, showing the relationships between different similarity measures (e.g., m, func, obj) and analogical relationships (i.e., anal.). We perform the widely-used Pearson correlation coefficient~\cite{pearson} and  Welch's t-test~\cite{welch1947generalization} to assess the statistical significance of the observed correlations. First, we could observe statistically-positive correlation of all similarities with the analogical relationships ($p << 0.05$) based on Welch’s t-test, implying that included similarities are helpful for inferring analogical relationship more or less. Second, each similarity score exhibits different correlation coefficients to the analogical relationship, implying a different importance of their role in inferring analogical relationship. Third, most similarity scores exhibit low correlations with others and only a few similarity scores exhibit high correlation (e.g., it v.s. iv). Overall, the statistical analysis indicates the potential benefits of different similarities to the analogical relationship inference; but at the same time there could be some redundant information among some similarities, indicating a potentially improving direction for the future work. }

\textcolor{black}{In summary, the current design choices (i.e., re-ranking step, KG embedding model, knowledge types, similarity types, and weights) all positively contribute to the effectiveness of \app.}


\begin{table}[]
		\centering
		\vspace{-2mm}
		\caption{Contribution of Different Similarities}
        \vspace{-4mm}
		\label{tab:rerank_sim}
		\footnotesize
		\begin{tabular}{|p{1.5cm}<{\centering}|c|c|c|p{0.7cm}<{\centering}|p{0.8cm}<{\centering}|}
			\hline
			Similarity     & MRR & Hit@1 & Hit@3 & Hit@5 & Hit@10  \\ \hline
			$Sim_m$*     &    0.382  &  0.263  &     0.449   &   0.522&     0.590                         \\ \hline
			$Sim_{func}$* &  0.297    &  0.194  & 0.343   &    0.410      & 0.502                        \\ \hline
			$Sim_{obj}$*   & 0.340   &  0.229 &   0.390  &   0.465      &    0.573                      \\ \hline
			$Sim_{iv}$*   &  0.383  &   0.271 &  0.447  &  0.520     &    0.592                       \\ \hline
			$Sim_{it}$*  &   0.370  &  0.251   &  0.437  &      0.510   &   0.592                     \\ \hline
			$Sim_{ot}$*   &   0.381  & 0.265   &  \textbf{0.451}   & \textbf{0.527}    &    0.580                     \\ \hline
			$Sim_{neig}$*     & \textbf{0.385}&    \textbf{0.273}  &   0.447  &  \textbf{0.527}  &  0.578                            \\ \hline
			All  & 0.384 &   0.267 &   0.449 &   \textbf{0.527}&    \textbf{0.594} \\ \hline
		\end{tabular}
  \vspace{-5mm}
	\end{table}

\subsection{RQ4: Scalability}
\label{sec:rq4}
In this RQ, we explore the scalability of \app. 

\parabf{\textcolor{black}{Online Cost.}}
\textcolor{black}{
The online inference time of KGE4AR is less than one second for one query in RQ1 and RQ2. It consists of two main steps: candidate API method retrieval and re-rank. The re-rank step's time is proportional to the number of candidates and remains constant once the candidates are determined. The retrieval step's time depends on the API KG size and the vector database used.
To address this, we employed the highly efficient vector index mechanism provided by Milvus, a scalable and highly available vector database. Milvus has been proven to achieve an average latency of milliseconds for vector search and retrieval on trillion-vector datasets~\cite{milvus,wang2021milvus}. This ensures that the retrieval step of KGE4AR is performed efficiently, even as the size of the API KG increases.}

\parabf{\textcolor{black}{Offline Cost.}}
\textcolor{black}{ We primarily discuss the offline costs of \app with different KG scales. Table~\ref{tab:scale} presents the construction costs for three API KGs: large-scale, medium-scale, and small-scale. The costs are calculated on a Linux server with a 36-core CPU and 128GB RAM. The columns \textit{Input}, \textit{Construct.}, and \textit{Embed.} represent the time for downloading/preparing documentations as input, API KG construction, and API KG embedding, respectively.}
Although the number of entities increases by 2,019 times from a small-scale API KG to a large-scale API KG, the time required for collecting inputs, API KG construction, and API KG embedding only increases by 386 times, 121 times, and 40 times, respectively.
Note that the KG construction and embedding are only executed once, and the KG could be incrementally extended when there are new libraries.

\begin{table}[]
	\centering
	\caption{\textcolor{black}{Offline Cost of \app at Different Scales}}
	\label{tab:scale}
	\footnotesize
	\vspace{-4mm}
	\begin{tabular}{|p{0.8cm}<{\centering}|c|c|c|p{0.8cm}<{\centering}|p{1.0cm}<{\centering}|p{0.7cm}<{\centering}|}
		\hline
		 Type & Library & Entity & Relation   & Input & Construct. & Embed. \\ \hline
		 Small    & 16&35K      &   2M        &     7m             &      49m          &       1.5h            \\ \hline
		 Medium  &    899& 2M     &   8M           &       1h           &     4h           &    6h           \\ \hline
		 Large &   35,773 & 72M     &   289M           &       45h           &     99h           &    60h          \\ \hline
	\end{tabular}
	\vspace{-7mm}
\end{table}

\textcolor{black}{
In summary, there is evidence to suggest that \app has the potential to scale effectively as the number of libraries increases.}


\subsection{Threats to Validity}
\label{sec:threats}

\textcolor{black}{\parabf{Internal Validity.}
A threat to the internal validity of our studies is the subjectivity of human annotations in RQ2. To mitigate this threat, we implemented measures such as multiple annotators, conflict resolution, and reporting agreement coefficients. These practices were employed to minimize bias and ensure the reliability of the human annotations.}

\parabf{\textcolor{black}{External Validity.}}
\textcolor{black}{
A limitation of our study is the exclusive focus on Java libraries, potentially limiting the generalizability of our findings to other programming languages. However, the core concept of our approach, involving the construction of a unified knowledge graph across libraries, remains applicable. While our knowledge graph design is not limited to Java, it can be extended to accommodate libraries from other object-oriented languages. However, specific implementation adjustments would be required. For example, supporting languages like Python, which lack strong typing, would necessitate modifying the schema. 
Future work will explore more programming languages for a comprehensive evaluation of our approach's effectiveness across diverse language environments.}


\parabf{\textcolor{black}{Construct Validity.}}
\textcolor{black}{A common threat is that the baselines we used in RQ1 and RQ2 are implemented by ourselves due to publicly unavailable implementations. 
However, we carefully reproduced and tested the baselines to avoid introducing errors.
Another threat is the way similarity weights are determined.
We tuned the weights through the benchmark in RQ1 and the weights may overfit the benchmark. 
To mitigate this threat, we tuned the weights on a small-scale API KG instead of the large-scale API KG used by RQ1. 
Figure~\ref{fig:fold_num} also shows that our weights do not overfit the benchmark.}






\section{Conclusions}
This work proposes \app, a novel documentation-based approach that leverages knowledge graph (KG) embedding to recommend analogical APIs during library migration. In particular, \app  proposes a novel unified API KG to comprehensively and structurally represent three types of knowledge in documentation, which could better capture the high-level semantics. In addition, \app then proposes to embed the unified API KG, which enables more effective and scalable similarity calculation. 
We implement \app as a fully automated technique with constructing a unified API KG for 35,773 Java libraries. We further evaluate \app in two API recommendation scenarios (i.e., with given target libraries or without given target libraries), and our results show that \app substantially outperforms state-of-the-art documentation-based techniques in both evaluation scenarios in terms of all metrics.  In addition, we further investigate the scalability of \app and find that \app can well scale with the increasing number of libraries.

\section{Data Availability}
All the data and code could be found in our replication package~\cite{replication_package}.

\section*{Acknowledgments}
This work is supported by National Natural Science Foundation of China under Grant No. 61972098.

\normalem
\bibliographystyle{ACM-Reference-Format}
\balance
\bibliography{citelist}


\begin{thebibliography}{79}


\ifx \showCODEN    \undefined \def \showCODEN     #1{\unskip}     \fi
\ifx \showDOI      \undefined \def \showDOI       #1{#1}\fi
\ifx \showISBNx    \undefined \def \showISBNx     #1{\unskip}     \fi
\ifx \showISBNxiii \undefined \def \showISBNxiii  #1{\unskip}     \fi
\ifx \showISSN     \undefined \def \showISSN      #1{\unskip}     \fi
\ifx \showLCCN     \undefined \def \showLCCN      #1{\unskip}     \fi
\ifx \shownote     \undefined \def \shownote      #1{#1}          \fi
\ifx \showarticletitle \undefined \def \showarticletitle #1{#1}   \fi
\ifx \showURL      \undefined \def \showURL       {\relax}        \fi
\providecommand\bibfield[2]{#2}
\providecommand\bibinfo[2]{#2}
\providecommand\natexlab[1]{#1}
\providecommand\showeprint[2][]{arXiv:#2}

\bibitem[Alr(2023)]%
        {Alrubayedataset}
 \bibinfo{year}{2023}\natexlab{}.
\newblock \bibinfo{booktitle}{\emph{Alrubaye et al. Dataset}}.
\newblock
\urldef\tempurl%
\url{http://migrationlab.net/ds/groundTruth_icpc2019.html}
\showURL{%
Retrieved January 20, 2023 from \tempurl}


\bibitem[com(2023)]%
        {commons-logging}
 \bibinfo{year}{2023}\natexlab{}.
\newblock \bibinfo{booktitle}{\emph{Apache Commons Logging}}.
\newblock
\urldef\tempurl%
\url{https://mvnrepository.com/artifact/commons-logging/commons-logging}
\showURL{%
Retrieved January 20, 2023 from \tempurl}


\bibitem[awe(2023)]%
        {awesomejava}
 \bibinfo{year}{2023}\natexlab{}.
\newblock \bibinfo{booktitle}{\emph{awesome-java}}.
\newblock
\urldef\tempurl%
\url{https://github.com/akullpp/awesome-java}
\showURL{%
Retrieved January 20, 2023 from \tempurl}


\bibitem[gso(2023)]%
        {gson}
 \bibinfo{year}{2023}\natexlab{}.
\newblock \bibinfo{booktitle}{\emph{com.google.code.gson:gson}}.
\newblock
\urldef\tempurl%
\url{https://mvnrepository.com/artifact/com.google.code.gson/gson}
\showURL{%
Retrieved January 20, 2023 from \tempurl}


\bibitem[doc(2023)]%
        {doccomment}
 \bibinfo{year}{2023}\natexlab{}.
\newblock \bibinfo{booktitle}{\emph{Doc Comment}}.
\newblock
\urldef\tempurl%
\url{https://www.oracle.com/technical-resources/articles/java/javadoc-tool.html}
\showURL{%
Retrieved January 20, 2023 from \tempurl}


\bibitem[dom(2023)]%
        {dom4j}
 \bibinfo{year}{2023}\natexlab{}.
\newblock \bibinfo{booktitle}{\emph{dom4j}}.
\newblock
\urldef\tempurl%
\url{https://github.com/dom4j/dom4j}
\showURL{%
Retrieved January 20, 2023 from \tempurl}


\bibitem[dsl(2023)]%
        {dsl-json}
 \bibinfo{year}{2023}\natexlab{}.
\newblock \bibinfo{booktitle}{\emph{dsl-json}}.
\newblock
\urldef\tempurl%
\url{https://github.com/ngs-doo/dsl-json}
\showURL{%
Retrieved January 20, 2023 from \tempurl}


\bibitem[ela(2023)]%
        {elasticsearch}
 \bibinfo{year}{2023}\natexlab{}.
\newblock \bibinfo{booktitle}{\emph{Elasticsearch}}.
\newblock
\urldef\tempurl%
\url{https://github.com/elastic/elasticsearch}
\showURL{%
Retrieved January 20, 2023 from \tempurl}


\bibitem[fun(2023a)]%
        {funcverbnet}
 \bibinfo{year}{2023}\natexlab{a}.
\newblock \bibinfo{booktitle}{\emph{FuncVerbNet}}.
\newblock
\urldef\tempurl%
\url{https://github.com/FudanSELab/funcverbnet}
\showURL{%
Retrieved January 20, 2023 from \tempurl}


\bibitem[jav(2023)]%
        {javaparser}
 \bibinfo{year}{2023}\natexlab{}.
\newblock \bibinfo{booktitle}{\emph{JavaParser}}.
\newblock
\urldef\tempurl%
\url{https://javaparser.org/}
\showURL{%
Retrieved January 20, 2023 from \tempurl}


\bibitem[jun(2023)]%
        {junit}
 \bibinfo{year}{2023}\natexlab{}.
\newblock \bibinfo{booktitle}{\emph{junit:junit}}.
\newblock
\urldef\tempurl%
\url{https://mvnrepository.com/artifact/junit/junit}
\showURL{%
Retrieved January 20, 2023 from \tempurl}


\bibitem[lib(2023)]%
        {libariesiodata}
 \bibinfo{year}{2023}\natexlab{}.
\newblock \bibinfo{booktitle}{\emph{Libaries.io open data}}.
\newblock
\urldef\tempurl%
\url{https://libraries.io/data}
\showURL{%
Retrieved January 20, 2023 from \tempurl}


\bibitem[mav(2023)]%
        {mavencenter}
 \bibinfo{year}{2023}\natexlab{}.
\newblock \bibinfo{booktitle}{\emph{{Maven} {Central} {Repository}}}.
\newblock
\urldef\tempurl%
\url{https://mvnrepository.com}
\showURL{%
Retrieved January 20, 2023 from \tempurl}


\bibitem[mil(2023)]%
        {milvus}
 \bibinfo{year}{2023}\natexlab{}.
\newblock \bibinfo{booktitle}{\emph{milvus}}.
\newblock
\urldef\tempurl%
\url{https://github.com/milvus-io/milvus}
\showURL{%
Retrieved January 20, 2023 from \tempurl}


\bibitem[org(2023a)]%
        {org.json}
 \bibinfo{year}{2023}\natexlab{a}.
\newblock \bibinfo{booktitle}{\emph{org.json:json}}.
\newblock
\urldef\tempurl%
\url{https://mvnrepository.com/artifact/org.json/json}
\showURL{%
Retrieved January 20, 2023 from \tempurl}


\bibitem[slf(2023)]%
        {slf4j}
 \bibinfo{year}{2023}\natexlab{}.
\newblock \bibinfo{booktitle}{\emph{org.slf4j:slf4j-api}}.
\newblock
\urldef\tempurl%
\url{https://mvnrepository.com/artifact/org.slf4j/slf4j-api}
\showURL{%
Retrieved January 20, 2023 from \tempurl}


\bibitem[org(2023b)]%
        {org.testing}
 \bibinfo{year}{2023}\natexlab{b}.
\newblock \bibinfo{booktitle}{\emph{org.testing:testing}}.
\newblock
\urldef\tempurl%
\url{https://mvnrepository.com/artifact/org.testng/testng}
\showURL{%
Retrieved January 20, 2023 from \tempurl}


\bibitem[Big(2023)]%
        {BigGraphGithub}
 \bibinfo{year}{2023}\natexlab{}.
\newblock \bibinfo{booktitle}{\emph{PyTorch-BigGraph}}.
\newblock
\urldef\tempurl%
\url{https://github.com/facebookresearch/PyTorch-BigGraph}
\showURL{%
Retrieved January 20, 2023 from \tempurl}


\bibitem[asc(2023)]%
        {asc_service}
 \bibinfo{year}{2023}\natexlab{}.
\newblock \bibinfo{booktitle}{\emph{{RAPIM} Service}}.
\newblock
\urldef\tempurl%
\url{http://migrationlab.net/MigrationWebService.php}
\showURL{%
Retrieved January 20, 2023 from \tempurl}


\bibitem[rep(2023)]%
        {replication_package}
 \bibinfo{year}{2023}\natexlab{}.
\newblock \bibinfo{booktitle}{\emph{Replication Package}}.
\newblock
\urldef\tempurl%
\url{https://github.com/FudanSELab/KGE4AR}
\showURL{%
Retrieved August 20, 2023 from \tempurl}


\bibitem[fun(2023b)]%
        {funcverbnet-replication}
 \bibinfo{year}{2023}\natexlab{b}.
\newblock \bibinfo{booktitle}{\emph{Replication Package of FuncVerbNet}}.
\newblock
\urldef\tempurl%
\url{https://github.com/FudanSELab/Research-FSE2020-FuncVerb}
\showURL{%
Retrieved January 20, 2023 from \tempurl}


\bibitem[spa(2023)]%
        {spacy}
 \bibinfo{year}{2023}\natexlab{}.
\newblock \bibinfo{booktitle}{\emph{Spacy}}.
\newblock
\urldef\tempurl%
\url{https://spacy.io/}
\showURL{%
Retrieved January 20, 2023 from \tempurl}


\bibitem[tes(2023)]%
        {testng}
 \bibinfo{year}{2023}\natexlab{}.
\newblock \bibinfo{booktitle}{\emph{{TestNG}}}.
\newblock
\urldef\tempurl%
\url{https://mvnrepository.com/artifact/org.testng/testng}
\showURL{%
Retrieved January 20, 2023 from \tempurl}


\bibitem[Tey(2023)]%
        {Teytondataset}
 \bibinfo{year}{2023}\natexlab{}.
\newblock \bibinfo{booktitle}{\emph{Teyton et al. Dataset}}.
\newblock
\urldef\tempurl%
\url{http://web.archive.org/web/20160412155706/http://www.labri.fr/perso/cteyton/Matching/lang\_commons\_guava.html}
\showURL{%
Retrieved January 20, 2023 from \tempurl}


\bibitem[zip(2023)]%
        {zipfile}
 \bibinfo{year}{2023}\natexlab{}.
\newblock \bibinfo{booktitle}{\emph{zipfile}}.
\newblock
\urldef\tempurl%
\url{https://docs.python.org/3/library/zipfile.html}
\showURL{%
Retrieved January 20, 2023 from \tempurl}


\bibitem[Abdi and Williams(2010)]%
        {pca}
\bibfield{author}{\bibinfo{person}{Herv{\'e} Abdi} {and}
  \bibinfo{person}{Lynne~J Williams}.} \bibinfo{year}{2010}\natexlab{}.
\newblock \showarticletitle{Principal component analysis}.
\newblock \bibinfo{journal}{\emph{Wiley interdisciplinary reviews:
  computational statistics}} \bibinfo{volume}{2}, \bibinfo{number}{4}
  (\bibinfo{year}{2010}), \bibinfo{pages}{433--459}.
\newblock


\bibitem[Alrubaye and Mkaouer(2018)]%
        {CASCON18funmap}
\bibfield{author}{\bibinfo{person}{Hussein Alrubaye} {and}
  \bibinfo{person}{Mohamed~Wiem Mkaouer}.} \bibinfo{year}{2018}\natexlab{}.
\newblock \showarticletitle{Automating the detection of third-party Java
  library migration at the function level}. In
  \bibinfo{booktitle}{\emph{Proceedings of the 28th Annual International
  Conference on Computer Science and Software Engineering, {CASCON} 2018,
  Markham, Ontario, Canada, October 29-31, 2018}}. \bibinfo{publisher}{{ACM}},
  \bibinfo{pages}{60--71}.
\newblock
\urldef\tempurl%
\url{https://dl.acm.org/citation.cfm?id=3291299}
\showURL{%
\tempurl}


\bibitem[Alrubaye et~al\mbox{.}(2020)]%
        {asc20apimigration}
\bibfield{author}{\bibinfo{person}{Hussein Alrubaye},
  \bibinfo{person}{Mohamed~Wiem Mkaouer}, \bibinfo{person}{Igor Khokhlov},
  \bibinfo{person}{Leon Reznik}, \bibinfo{person}{Ali Ouni}, {and}
  \bibinfo{person}{Jason Mcgoff}.} \bibinfo{year}{2020}\natexlab{}.
\newblock \showarticletitle{Learning to recommend third-party library migration
  opportunities at the {API} level}.
\newblock \bibinfo{journal}{\emph{Appl. Soft Comput.}}  \bibinfo{volume}{90}
  (\bibinfo{year}{2020}), \bibinfo{pages}{106140}.
\newblock
\urldef\tempurl%
\url{https://doi.org/10.1016/j.asoc.2020.106140}
\showDOI{\tempurl}


\bibitem[Alrubaye et~al\mbox{.}(2019a)]%
        {icsme19apimig}
\bibfield{author}{\bibinfo{person}{Hussein Alrubaye},
  \bibinfo{person}{Mohamed~Wiem Mkaouer}, {and} \bibinfo{person}{Ali Ouni}.}
  \bibinfo{year}{2019}\natexlab{a}.
\newblock \showarticletitle{MigrationMiner: An Automated Detection Tool of
  Third-Party Java Library Migration at the Method Level}. In
  \bibinfo{booktitle}{\emph{2019 {IEEE} International Conference on Software
  Maintenance and Evolution, {ICSME} 2019, Cleveland, OH, USA, September 29 -
  October 4, 2019}}. \bibinfo{publisher}{{IEEE}}, \bibinfo{pages}{414--417}.
\newblock
\urldef\tempurl%
\url{https://doi.org/10.1109/ICSME.2019.00072}
\showDOI{\tempurl}


\bibitem[Alrubaye et~al\mbox{.}(2019b)]%
        {icpc19apimap}
\bibfield{author}{\bibinfo{person}{Hussein Alrubaye},
  \bibinfo{person}{Mohamed~Wiem Mkaouer}, {and} \bibinfo{person}{Ali Ouni}.}
  \bibinfo{year}{2019}\natexlab{b}.
\newblock \showarticletitle{On the use of information retrieval to automate the
  detection of third-party Java library migration at the method level}. In
  \bibinfo{booktitle}{\emph{Proceedings of the 27th International Conference on
  Program Comprehension, {ICPC} 2019, Montreal, QC, Canada, May 25-31, 2019}}.
  \bibinfo{publisher}{{IEEE} / {ACM}}, \bibinfo{pages}{347--357}.
\newblock
\urldef\tempurl%
\url{https://doi.org/10.1109/ICPC.2019.00053}
\showDOI{\tempurl}


\bibitem[Alrubaye et~al\mbox{.}(2019c)]%
        {sevis19libevolution}
\bibfield{author}{\bibinfo{person}{Hussein Alrubaye},
  \bibinfo{person}{Mohamed~Wiem Mkaouer}, {and} \bibinfo{person}{Anthony
  Peruma}.} \bibinfo{year}{2019}\natexlab{c}.
\newblock \showarticletitle{Variability in Library Evolution}.
\newblock In \bibinfo{booktitle}{\emph{Software Engineering for Variability
  Intensive Systems - Foundations and Applications}}.
  \bibinfo{publisher}{Auerbach Publications / Taylor {\&} Francis},
  \bibinfo{pages}{295--320}.
\newblock
\urldef\tempurl%
\url{https://doi.org/10.1201/9780429022067-13}
\showDOI{\tempurl}


\bibitem[Bordes et~al\mbox{.}(2013)]%
        {transE}
\bibfield{author}{\bibinfo{person}{Antoine Bordes}, \bibinfo{person}{Nicolas
  Usunier}, \bibinfo{person}{Alberto Garc{\'{\i}}a{-}Dur{\'{a}}n},
  \bibinfo{person}{Jason Weston}, {and} \bibinfo{person}{Oksana Yakhnenko}.}
  \bibinfo{year}{2013}\natexlab{}.
\newblock \showarticletitle{Translating Embeddings for Modeling
  Multi-relational Data}. In \bibinfo{booktitle}{\emph{Advances in Neural
  Information Processing Systems 26: 27th Annual Conference on Neural
  Information Processing Systems 2013. Proceedings of a meeting held December
  5-8, 2013, Lake Tahoe, Nevada, United States}}. \bibinfo{pages}{2787--2795}.
\newblock
\urldef\tempurl%
\url{https://proceedings.neurips.cc/paper/2013/hash/1cecc7a77928ca8133fa24680a88d2f9-Abstract.html}
\showURL{%
\tempurl}


\bibitem[Chen et~al\mbox{.}(2016)]%
        {saner16libmigration}
\bibfield{author}{\bibinfo{person}{Chunyang Chen}, \bibinfo{person}{Sa Gao},
  {and} \bibinfo{person}{Zhenchang Xing}.} \bibinfo{year}{2016}\natexlab{}.
\newblock \showarticletitle{Mining Analogical Libraries in Q{\&}A Discussions -
  Incorporating Relational and Categorical Knowledge into Word Embedding}. In
  \bibinfo{booktitle}{\emph{{IEEE} 23rd International Conference on Software
  Analysis, Evolution, and Reengineering, {SANER} 2016, Suita, Osaka, Japan,
  March 14-18, 2016 - Volume 1}}. \bibinfo{publisher}{{IEEE} Computer Society},
  \bibinfo{pages}{338--348}.
\newblock
\urldef\tempurl%
\url{https://doi.org/10.1109/SANER.2016.21}
\showDOI{\tempurl}


\bibitem[Chen et~al\mbox{.}(2021)]%
        {tse19apimigration}
\bibfield{author}{\bibinfo{person}{Chunyang Chen}, \bibinfo{person}{Zhenchang
  Xing}, \bibinfo{person}{Yang Liu}, {and} \bibinfo{person}{Kent Ong~Long
  Xiong}.} \bibinfo{year}{2021}\natexlab{}.
\newblock \showarticletitle{Mining Likely Analogical APIs Across Third-Party
  Libraries via Large-Scale Unsupervised {API} Semantics Embedding}.
\newblock \bibinfo{journal}{\emph{{IEEE} Trans. Software Eng.}}
  \bibinfo{volume}{47}, \bibinfo{number}{3} (\bibinfo{year}{2021}),
  \bibinfo{pages}{432--447}.
\newblock
\urldef\tempurl%
\url{https://doi.org/10.1109/TSE.2019.2896123}
\showDOI{\tempurl}


\bibitem[Coelho and Valente(2017)]%
        {fse17projectfail}
\bibfield{author}{\bibinfo{person}{Jailton Coelho} {and}
  \bibinfo{person}{Marco~T{\'{u}}lio Valente}.}
  \bibinfo{year}{2017}\natexlab{}.
\newblock \showarticletitle{Why modern open source projects fail}. In
  \bibinfo{booktitle}{\emph{Proceedings of the 2017 11th Joint Meeting on
  Foundations of Software Engineering, {ESEC/FSE} 2017, Paderborn, Germany,
  September 4-8, 2017}}. \bibinfo{publisher}{{ACM}}, \bibinfo{pages}{186--196}.
\newblock
\urldef\tempurl%
\url{https://doi.org/10.1145/3106237.3106246}
\showDOI{\tempurl}


\bibitem[Cohen et~al\mbox{.}(2009)]%
        {pearson}
\bibfield{author}{\bibinfo{person}{Israel Cohen}, \bibinfo{person}{Yiteng
  Huang}, \bibinfo{person}{Jingdong Chen}, \bibinfo{person}{Jacob Benesty},
  \bibinfo{person}{Jacob Benesty}, \bibinfo{person}{Jingdong Chen},
  \bibinfo{person}{Yiteng Huang}, {and} \bibinfo{person}{Israel Cohen}.}
  \bibinfo{year}{2009}\natexlab{}.
\newblock \showarticletitle{Pearson correlation coefficient}.
\newblock \bibinfo{journal}{\emph{Noise reduction in speech processing}}
  (\bibinfo{year}{2009}), \bibinfo{pages}{1--4}.
\newblock


\bibitem[Cossette and Walker(2012)]%
        {fse12libmigration}
\bibfield{author}{\bibinfo{person}{Bradley Cossette} {and}
  \bibinfo{person}{Robert~J. Walker}.} \bibinfo{year}{2012}\natexlab{}.
\newblock \showarticletitle{Seeking the ground truth: a retroactive study on
  the evolution and migration of software libraries}. In
  \bibinfo{booktitle}{\emph{20th {ACM} {SIGSOFT} Symposium on the Foundations
  of Software Engineering (FSE-20), SIGSOFT/FSE'12, Cary, NC, {USA} - November
  11 - 16, 2012}}. \bibinfo{publisher}{{ACM}}, \bibinfo{pages}{55}.
\newblock
\urldef\tempurl%
\url{https://doi.org/10.1145/2393596.2393661}
\showDOI{\tempurl}


\bibitem[Du et~al\mbox{.}({[n.\,d.]})]%
        {kglsurvey}
\bibfield{author}{\bibinfo{person}{Xueying Du}, \bibinfo{person}{Mingwei Liu},
  \bibinfo{person}{Liwei Shen}, {and} \bibinfo{person}{Xin Peng}.}
  \bibinfo{year}{[n.\,d.]}\natexlab{}.
\newblock \showarticletitle{Research on Knowledge Graph Representation Learning
  Methods for Link Prediction: A Review}.
\newblock \bibinfo{journal}{\emph{Journal of Software}}
  (\bibinfo{year}{[n.\,d.]}).
\newblock


\bibitem[Gao et~al\mbox{.}(2020)]%
        {beamsearch}
\bibfield{author}{\bibinfo{person}{Zhipeng Gao}, \bibinfo{person}{Xin Xia},
  \bibinfo{person}{John Grundy}, \bibinfo{person}{David Lo}, {and}
  \bibinfo{person}{Yuan{-}Fang Li}.} \bibinfo{year}{2020}\natexlab{}.
\newblock \showarticletitle{Generating Question Titles for Stack Overflow from
  Mined Code Snippets}.
\newblock \bibinfo{journal}{\emph{{ACM} Trans. Softw. Eng. Methodol.}}
  \bibinfo{volume}{29}, \bibinfo{number}{4} (\bibinfo{year}{2020}),
  \bibinfo{pages}{26:1--26:37}.
\newblock
\urldef\tempurl%
\url{https://doi.org/10.1145/3401026}
\showDOI{\tempurl}


\bibitem[Germ{\'{a}}n and Penta(2012)]%
        {softw12license}
\bibfield{author}{\bibinfo{person}{Daniel~M. Germ{\'{a}}n} {and}
  \bibinfo{person}{Massimiliano~Di Penta}.} \bibinfo{year}{2012}\natexlab{}.
\newblock \showarticletitle{A Method for Open Source License Compliance of Java
  Applications}.
\newblock \bibinfo{journal}{\emph{{IEEE} Softw.}} \bibinfo{volume}{29},
  \bibinfo{number}{3} (\bibinfo{year}{2012}), \bibinfo{pages}{58--63}.
\newblock
\urldef\tempurl%
\url{https://doi.org/10.1109/MS.2012.50}
\showDOI{\tempurl}


\bibitem[He et~al\mbox{.}(2021a)]%
        {fse21libmigration}
\bibfield{author}{\bibinfo{person}{Hao He}, \bibinfo{person}{Runzhi He},
  \bibinfo{person}{Haiqiao Gu}, {and} \bibinfo{person}{Minghui Zhou}.}
  \bibinfo{year}{2021}\natexlab{a}.
\newblock \showarticletitle{A large-scale empirical study on Java library
  migrations: prevalence, trends, and rationales}. In
  \bibinfo{booktitle}{\emph{{ESEC/FSE} '21: 29th {ACM} Joint European Software
  Engineering Conference and Symposium on the Foundations of Software
  Engineering, Athens, Greece, August 23-28, 2021}}.
  \bibinfo{publisher}{{ACM}}, \bibinfo{pages}{478--490}.
\newblock
\urldef\tempurl%
\url{https://doi.org/10.1145/3468264.3468571}
\showDOI{\tempurl}


\bibitem[He et~al\mbox{.}(2021b)]%
        {saner21libmigration}
\bibfield{author}{\bibinfo{person}{Hao He}, \bibinfo{person}{Yulin Xu},
  \bibinfo{person}{Yixiao Ma}, \bibinfo{person}{Yifei Xu},
  \bibinfo{person}{Guangtai Liang}, {and} \bibinfo{person}{Minghui Zhou}.}
  \bibinfo{year}{2021}\natexlab{b}.
\newblock \showarticletitle{A Multi-Metric Ranking Approach for Library
  Migration Recommendations}. In \bibinfo{booktitle}{\emph{28th {IEEE}
  International Conference on Software Analysis, Evolution and Reengineering,
  {SANER} 2021, Honolulu, HI, USA, March 9-12, 2021}}.
  \bibinfo{publisher}{{IEEE}}, \bibinfo{pages}{72--83}.
\newblock
\urldef\tempurl%
\url{https://doi.org/10.1109/SANER50967.2021.00016}
\showDOI{\tempurl}


\bibitem[Kula et~al\mbox{.}(2018)]%
        {ese18libmigration}
\bibfield{author}{\bibinfo{person}{Raula~Gaikovina Kula},
  \bibinfo{person}{Daniel~M. Germ{\'{a}}n}, \bibinfo{person}{Ali Ouni},
  \bibinfo{person}{Takashi Ishio}, {and} \bibinfo{person}{Katsuro Inoue}.}
  \bibinfo{year}{2018}\natexlab{}.
\newblock \showarticletitle{Do developers update their library dependencies? -
  An empirical study on the impact of security advisories on library
  migration}.
\newblock \bibinfo{journal}{\emph{Empir. Softw. Eng.}} \bibinfo{volume}{23},
  \bibinfo{number}{1} (\bibinfo{year}{2018}), \bibinfo{pages}{384--417}.
\newblock
\urldef\tempurl%
\url{https://doi.org/10.1007/s10664-017-9521-5}
\showDOI{\tempurl}


\bibitem[Lerer et~al\mbox{.}(2019)]%
        {pbg}
\bibfield{author}{\bibinfo{person}{Adam Lerer}, \bibinfo{person}{Ledell Wu},
  \bibinfo{person}{Jiajun Shen}, \bibinfo{person}{Timoth{\'{e}}e Lacroix},
  \bibinfo{person}{Luca Wehrstedt}, \bibinfo{person}{Abhijit Bose}, {and}
  \bibinfo{person}{Alex Peysakhovich}.} \bibinfo{year}{2019}\natexlab{}.
\newblock \showarticletitle{Pytorch-BigGraph: {A} Large Scale Graph Embedding
  System}. In \bibinfo{booktitle}{\emph{Proceedings of Machine Learning and
  Systems 2019, MLSys 2019, Stanford, CA, USA, March 31 - April 2, 2019}}.
  \bibinfo{publisher}{mlsys.org}.
\newblock
\urldef\tempurl%
\url{https://proceedings.mlsys.org/book/282.pdf}
\showURL{%
\tempurl}


\bibitem[Li et~al\mbox{.}(2018)]%
        {icsme2018apicaveat}
\bibfield{author}{\bibinfo{person}{Hongwei Li}, \bibinfo{person}{Sirui Li},
  \bibinfo{person}{Jiamou Sun}, \bibinfo{person}{Zhenchang Xing},
  \bibinfo{person}{Xin Peng}, \bibinfo{person}{Mingwei Liu}, {and}
  \bibinfo{person}{Xuejiao Zhao}.} \bibinfo{year}{2018}\natexlab{}.
\newblock \showarticletitle{Improving {API} Caveats Accessibility by Mining
  {API} Caveats Knowledge Graph}. In \bibinfo{booktitle}{\emph{2018 {IEEE}
  International Conference on Software Maintenance and Evolution, {ICSME} 2018,
  Madrid, Spain, September 23-29, 2018}}. \bibinfo{publisher}{{IEEE} Computer
  Society}, \bibinfo{pages}{183--193}.
\newblock
\urldef\tempurl%
\url{https://doi.org/10.1109/ICSME.2018.00028}
\showDOI{\tempurl}


\bibitem[Lim(1994)]%
        {softw94libreuse}
\bibfield{author}{\bibinfo{person}{Wayne~C. Lim}.}
  \bibinfo{year}{1994}\natexlab{}.
\newblock \showarticletitle{Effects of Reuse on Quality, Productivity, and
  Economics}.
\newblock \bibinfo{journal}{\emph{{IEEE} Softw.}} \bibinfo{volume}{11},
  \bibinfo{number}{5} (\bibinfo{year}{1994}), \bibinfo{pages}{23--30}.
\newblock
\urldef\tempurl%
\url{https://doi.org/10.1109/52.311048}
\showDOI{\tempurl}


\bibitem[Lin et~al\mbox{.}(2015)]%
        {transr}
\bibfield{author}{\bibinfo{person}{Yankai Lin}, \bibinfo{person}{Zhiyuan Liu},
  \bibinfo{person}{Maosong Sun}, \bibinfo{person}{Yang Liu}, {and}
  \bibinfo{person}{Xuan Zhu}.} \bibinfo{year}{2015}\natexlab{}.
\newblock \showarticletitle{Learning Entity and Relation Embeddings for
  Knowledge Graph Completion}. In \bibinfo{booktitle}{\emph{Proceedings of the
  Twenty-Ninth {AAAI} Conference on Artificial Intelligence, January 25-30,
  2015, Austin, Texas, {USA}}}. \bibinfo{publisher}{{AAAI} Press},
  \bibinfo{pages}{2181--2187}.
\newblock
\urldef\tempurl%
\url{http://www.aaai.org/ocs/index.php/AAAI/AAAI15/paper/view/9571}
\showURL{%
\tempurl}


\bibitem[Liu et~al\mbox{.}(2022a)]%
        {fse22taskkg}
\bibfield{author}{\bibinfo{person}{Mingwei Liu}, \bibinfo{person}{Xin Peng},
  \bibinfo{person}{Andrian Marcus}, \bibinfo{person}{Christoph Treude},
  \bibinfo{person}{Jiazhan Xie}, \bibinfo{person}{Huanjun Xu}, {and}
  \bibinfo{person}{Yanjun Yang}.} \bibinfo{year}{2022}\natexlab{a}.
\newblock \showarticletitle{How to formulate specific how-to questions in
  software development?}. In \bibinfo{booktitle}{\emph{Proceedings of the 30th
  {ACM} Joint European Software Engineering Conference and Symposium on the
  Foundations of Software Engineering, {ESEC/FSE} 2022, Singapore, Singapore,
  November 14-18, 2022}}. \bibinfo{publisher}{{ACM}},
  \bibinfo{pages}{306--318}.
\newblock
\urldef\tempurl%
\url{https://doi.org/10.1145/3540250.3549160}
\showDOI{\tempurl}


\bibitem[Liu et~al\mbox{.}(2022b)]%
        {tse21developerneed}
\bibfield{author}{\bibinfo{person}{Mingwei Liu}, \bibinfo{person}{Xin Peng},
  \bibinfo{person}{Andrian Marcus}, \bibinfo{person}{Shuangshuang Xing},
  \bibinfo{person}{Christoph Treude}, {and} \bibinfo{person}{Chengyuan Zhao}.}
  \bibinfo{year}{2022}\natexlab{b}.
\newblock \showarticletitle{API-Related Developer Information Needs in Stack
  Overflow}.
\newblock \bibinfo{journal}{\emph{{IEEE} Trans. Software Eng.}}
  \bibinfo{volume}{48}, \bibinfo{number}{11} (\bibinfo{year}{2022}),
  \bibinfo{pages}{4485--4500}.
\newblock
\urldef\tempurl%
\url{https://doi.org/10.1109/TSE.2021.3120203}
\showDOI{\tempurl}


\bibitem[Liu et~al\mbox{.}(2019)]%
        {fse19apisummary}
\bibfield{author}{\bibinfo{person}{Mingwei Liu}, \bibinfo{person}{Xin Peng},
  \bibinfo{person}{Andrian Marcus}, \bibinfo{person}{Zhenchang Xing},
  \bibinfo{person}{Wenkai Xie}, \bibinfo{person}{Shuangshuang Xing}, {and}
  \bibinfo{person}{Yang Liu}.} \bibinfo{year}{2019}\natexlab{}.
\newblock \showarticletitle{Generating query-specific class {API} summaries}.
  In \bibinfo{booktitle}{\emph{Proceedings of the {ACM} Joint Meeting on
  European Software Engineering Conference and Symposium on the Foundations of
  Software Engineering, {ESEC/SIGSOFT} {FSE} 2019, Tallinn, Estonia, August
  26-30, 2019}}. \bibinfo{publisher}{{ACM}}, \bibinfo{pages}{120--130}.
\newblock
\urldef\tempurl%
\url{https://doi.org/10.1145/3338906.3338971}
\showDOI{\tempurl}


\bibitem[Liu et~al\mbox{.}(2020b)]%
        {icsme20docgen}
\bibfield{author}{\bibinfo{person}{Mingwei Liu}, \bibinfo{person}{Xin Peng},
  \bibinfo{person}{Xiujie Meng}, \bibinfo{person}{Huanjun Xu},
  \bibinfo{person}{Shuangshuang Xing}, \bibinfo{person}{Xin Wang},
  \bibinfo{person}{Yang Liu}, {and} \bibinfo{person}{Gang Lv}.}
  \bibinfo{year}{2020}\natexlab{b}.
\newblock \showarticletitle{Source Code based On-demand Class Documentation
  Generation}. In \bibinfo{booktitle}{\emph{{IEEE} International Conference on
  Software Maintenance and Evolution, {ICSME} 2020, Adelaide, Australia,
  September 28 - October 2, 2020}}. \bibinfo{publisher}{{IEEE}},
  \bibinfo{pages}{864--865}.
\newblock
\urldef\tempurl%
\url{https://doi.org/10.1109/ICSME46990.2020.00114}
\showDOI{\tempurl}


\bibitem[Liu et~al\mbox{.}(2023)]%
        {tse2023aitaskmodelkg}
\bibfield{author}{\bibinfo{person}{Mingwei Liu}, \bibinfo{person}{Chengyuan
  Zhao}, \bibinfo{person}{Xin Peng}, \bibinfo{person}{Siming Yu},
  \bibinfo{person}{Haofen Wang}, {and} \bibinfo{person}{Chaofeng Sha}.}
  \bibinfo{year}{2023}\natexlab{}.
\newblock \showarticletitle{Task-Oriented ML/DL Library Recommendation based on
  a Knowledge Graph}.
\newblock \bibinfo{journal}{\emph{IEEE Transactions on Software Engineering}}
  (\bibinfo{year}{2023}).
\newblock


\bibitem[Liu et~al\mbox{.}(2020a)]%
        {ase20apicomp}
\bibfield{author}{\bibinfo{person}{Yang Liu}, \bibinfo{person}{Mingwei Liu},
  \bibinfo{person}{Xin Peng}, \bibinfo{person}{Christoph Treude},
  \bibinfo{person}{Zhenchang Xing}, {and} \bibinfo{person}{Xiaoxin Zhang}.}
  \bibinfo{year}{2020}\natexlab{a}.
\newblock \showarticletitle{Generating Concept based {API} Element Comparison
  Using a Knowledge Graph}. In \bibinfo{booktitle}{\emph{35th {IEEE/ACM}
  International Conference on Automated Software Engineering, {ASE} 2020,
  Melbourne, Australia, September 21-25, 2020}}. \bibinfo{publisher}{{IEEE}},
  \bibinfo{pages}{834--845}.
\newblock
\urldef\tempurl%
\url{https://doi.org/10.1145/3324884.3416628}
\showDOI{\tempurl}


\bibitem[Lu et~al\mbox{.}(2017)]%
        {ksem17apimap}
\bibfield{author}{\bibinfo{person}{Yangyang Lu}, \bibinfo{person}{Ge Li},
  \bibinfo{person}{Zelong Zhao}, \bibinfo{person}{Linfeng Wen}, {and}
  \bibinfo{person}{Zhi Jin}.} \bibinfo{year}{2017}\natexlab{}.
\newblock \showarticletitle{Learning to Infer {API} Mappings from {API}
  Documents}. In \bibinfo{booktitle}{\emph{Knowledge Science, Engineering and
  Management - 10th International Conference, {KSEM} 2017, Melbourne, VIC,
  Australia, August 19-20, 2017, Proceedings}} \emph{(\bibinfo{series}{Lecture
  Notes in Computer Science}, Vol.~\bibinfo{volume}{10412})}.
  \bibinfo{publisher}{Springer}, \bibinfo{pages}{237--248}.
\newblock
\urldef\tempurl%
\url{https://doi.org/10.1007/978-3-319-63558-3\_20}
\showDOI{\tempurl}


\bibitem[McHugh(2012)]%
        {kappa}
\bibfield{author}{\bibinfo{person}{Mary~L McHugh}.}
  \bibinfo{year}{2012}\natexlab{}.
\newblock \showarticletitle{Interrater reliability: the kappa statistic}.
\newblock \bibinfo{journal}{\emph{Biochemia Medica: Biochemia Medica}}
  \bibinfo{volume}{22}, \bibinfo{number}{3} (\bibinfo{year}{2012}),
  \bibinfo{pages}{276--282}.
\newblock


\bibitem[Miller(1995)]%
        {wordnet}
\bibfield{author}{\bibinfo{person}{George~A. Miller}.}
  \bibinfo{year}{1995}\natexlab{}.
\newblock \showarticletitle{WordNet: {A} Lexical Database for English}.
\newblock \bibinfo{journal}{\emph{Commun. {ACM}}} \bibinfo{volume}{38},
  \bibinfo{number}{11} (\bibinfo{year}{1995}), \bibinfo{pages}{39--41}.
\newblock
\urldef\tempurl%
\url{https://doi.org/10.1145/219717.219748}
\showDOI{\tempurl}


\bibitem[Mohagheghi and Conradi(2007)]%
        {ese07libreuse}
\bibfield{author}{\bibinfo{person}{Parastoo Mohagheghi} {and}
  \bibinfo{person}{Reidar Conradi}.} \bibinfo{year}{2007}\natexlab{}.
\newblock \showarticletitle{Quality, productivity and economic benefits of
  software reuse: a review of industrial studies}.
\newblock \bibinfo{journal}{\emph{Empir. Softw. Eng.}} \bibinfo{volume}{12},
  \bibinfo{number}{5} (\bibinfo{year}{2007}), \bibinfo{pages}{471--516}.
\newblock
\urldef\tempurl%
\url{https://doi.org/10.1007/s10664-007-9040-x}
\showDOI{\tempurl}


\bibitem[Nguyen et~al\mbox{.}(2014)]%
        {ase14apimap}
\bibfield{author}{\bibinfo{person}{Anh~Tuan Nguyen}, \bibinfo{person}{Hoan~Anh
  Nguyen}, \bibinfo{person}{Tung~Thanh Nguyen}, {and} \bibinfo{person}{Tien~N.
  Nguyen}.} \bibinfo{year}{2014}\natexlab{}.
\newblock \showarticletitle{Statistical learning approach for mining {API}
  usage mappings for code migration}. In \bibinfo{booktitle}{\emph{{ACM/IEEE}
  International Conference on Automated Software Engineering, {ASE} '14,
  Vasteras, Sweden - September 15 - 19, 2014}}. \bibinfo{publisher}{{ACM}},
  \bibinfo{pages}{457--468}.
\newblock
\urldef\tempurl%
\url{https://doi.org/10.1145/2642937.2643010}
\showDOI{\tempurl}


\bibitem[Pandita et~al\mbox{.}(2017)]%
        {jsep17tmap}
\bibfield{author}{\bibinfo{person}{Rahul Pandita}, \bibinfo{person}{Raoul
  Jetley}, \bibinfo{person}{Sithu~D. Sudarsan}, \bibinfo{person}{Tim Menzies},
  {and} \bibinfo{person}{Laurie~A. Williams}.} \bibinfo{year}{2017}\natexlab{}.
\newblock \showarticletitle{{TMAP:} Discovering relevant {API} methods through
  text mining of {API} documentation}.
\newblock \bibinfo{journal}{\emph{J. Softw. Evol. Process.}}
  \bibinfo{volume}{29}, \bibinfo{number}{12} (\bibinfo{year}{2017}).
\newblock
\urldef\tempurl%
\url{https://doi.org/10.1002/smr.1845}
\showDOI{\tempurl}


\bibitem[Pandita et~al\mbox{.}(2015)]%
        {scam15apimap}
\bibfield{author}{\bibinfo{person}{Rahul Pandita},
  \bibinfo{person}{Raoul~Praful Jetley}, \bibinfo{person}{Sithu~D. Sudarsan},
  {and} \bibinfo{person}{Laurie~A. Williams}.} \bibinfo{year}{2015}\natexlab{}.
\newblock \showarticletitle{Discovering likely mappings between APIs using text
  mining}. In \bibinfo{booktitle}{\emph{15th {IEEE} International Working
  Conference on Source Code Analysis and Manipulation, {SCAM} 2015, Bremen,
  Germany, September 27-28, 2015}}. \bibinfo{publisher}{{IEEE} Computer
  Society}, \bibinfo{pages}{231--240}.
\newblock
\urldef\tempurl%
\url{https://doi.org/10.1109/SCAM.2015.7335419}
\showDOI{\tempurl}


\bibitem[Peng et~al\mbox{.}(2018)]%
        {icsme18docgen}
\bibfield{author}{\bibinfo{person}{Xin Peng}, \bibinfo{person}{Yifan Zhao},
  \bibinfo{person}{Mingwei Liu}, \bibinfo{person}{Fengyi Zhang},
  \bibinfo{person}{Yang Liu}, \bibinfo{person}{Xin Wang}, {and}
  \bibinfo{person}{Zhenchang Xing}.} \bibinfo{year}{2018}\natexlab{}.
\newblock \showarticletitle{Automatic Generation of {API} Documentations for
  Open-Source Projects}. In \bibinfo{booktitle}{\emph{{IEEE} Third
  International Workshop on Dynamic Software Documentation, DySDoc@ICSME 2018,
  Madrid, Spain, September 25, 2018}}. \bibinfo{publisher}{{IEEE}},
  \bibinfo{pages}{7--8}.
\newblock
\urldef\tempurl%
\url{https://doi.org/10.1109/DySDoc3.2018.00010}
\showDOI{\tempurl}


\bibitem[Ren et~al\mbox{.}(2020)]%
        {ase2020apimiuse}
\bibfield{author}{\bibinfo{person}{Xiaoxue Ren}, \bibinfo{person}{Xinyuan Ye},
  \bibinfo{person}{Zhenchang Xing}, \bibinfo{person}{Xin Xia},
  \bibinfo{person}{Xiwei Xu}, \bibinfo{person}{Liming Zhu}, {and}
  \bibinfo{person}{Jianling Sun}.} \bibinfo{year}{2020}\natexlab{}.
\newblock \showarticletitle{API-Misuse Detection Driven by Fine-Grained
  API-Constraint Knowledge Graph}. In \bibinfo{booktitle}{\emph{35th {IEEE/ACM}
  International Conference on Automated Software Engineering, {ASE} 2020,
  Melbourne, Australia, September 21-25, 2020}}. \bibinfo{publisher}{{IEEE}},
  \bibinfo{pages}{461--472}.
\newblock
\urldef\tempurl%
\url{https://doi.org/10.1145/3324884.3416551}
\showDOI{\tempurl}


\bibitem[Robertson and Walker(1994)]%
        {bm25}
\bibfield{author}{\bibinfo{person}{Stephen~E. Robertson} {and}
  \bibinfo{person}{Steve Walker}.} \bibinfo{year}{1994}\natexlab{}.
\newblock \showarticletitle{Some Simple Effective Approximations to the
  2-Poisson Model for Probabilistic Weighted Retrieval}. In
  \bibinfo{booktitle}{\emph{Proceedings of the 17th Annual International
  {ACM-SIGIR} Conference on Research and Development in Information Retrieval.
  Dublin, Ireland, 3-6 July 1994 (Special Issue of the {SIGIR} Forum)}},
  \bibfield{editor}{\bibinfo{person}{W.~Bruce Croft} {and}
  \bibinfo{person}{C.~J. van Rijsbergen}} (Eds.).
  \bibinfo{publisher}{ACM/Springer}, \bibinfo{pages}{232--241}.
\newblock
\urldef\tempurl%
\url{https://doi.org/10.1007/978-1-4471-2099-5\_24}
\showDOI{\tempurl}


\bibitem[Su et~al\mbox{.}(2021)]%
        {su2021reducing}
\bibfield{author}{\bibinfo{person}{Yanqi Su}, \bibinfo{person}{Zhenchang Xing},
  \bibinfo{person}{Xin Peng}, \bibinfo{person}{Xin Xia}, \bibinfo{person}{Chong
  Wang}, \bibinfo{person}{Xiwei Xu}, {and} \bibinfo{person}{Liming Zhu}.}
  \bibinfo{year}{2021}\natexlab{}.
\newblock \showarticletitle{Reducing Bug Triaging Confusion by Learning from
  Mistakes with a Bug Tossing Knowledge Graph}. In
  \bibinfo{booktitle}{\emph{36th {IEEE/ACM} International Conference on
  Automated Software Engineering, {ASE} 2021, Melbourne, Australia, November
  15-19, 2021}}. \bibinfo{publisher}{{IEEE}}, \bibinfo{pages}{191--202}.
\newblock
\urldef\tempurl%
\url{https://doi.org/10.1109/ASE51524.2021.9678574}
\showDOI{\tempurl}


\bibitem[Teyton et~al\mbox{.}(2013)]%
        {wcre13funcmap}
\bibfield{author}{\bibinfo{person}{C{\'{e}}dric Teyton},
  \bibinfo{person}{Jean{-}R{\'{e}}my Falleri}, {and} \bibinfo{person}{Xavier
  Blanc}.} \bibinfo{year}{2013}\natexlab{}.
\newblock \showarticletitle{Automatic discovery of function mappings between
  similar libraries}. In \bibinfo{booktitle}{\emph{20th Working Conference on
  Reverse Engineering, {WCRE} 2013, Koblenz, Germany, October 14-17, 2013}}.
  \bibinfo{publisher}{{IEEE} Computer Society}, \bibinfo{pages}{192--201}.
\newblock
\urldef\tempurl%
\url{https://doi.org/10.1109/WCRE.2013.6671294}
\showDOI{\tempurl}


\bibitem[Trouillon et~al\mbox{.}(2016)]%
        {complex}
\bibfield{author}{\bibinfo{person}{Th{\'{e}}o Trouillon},
  \bibinfo{person}{Johannes Welbl}, \bibinfo{person}{Sebastian Riedel},
  \bibinfo{person}{{\'{E}}ric Gaussier}, {and} \bibinfo{person}{Guillaume
  Bouchard}.} \bibinfo{year}{2016}\natexlab{}.
\newblock \showarticletitle{Complex Embeddings for Simple Link Prediction}. In
  \bibinfo{booktitle}{\emph{Proceedings of the 33nd International Conference on
  Machine Learning, {ICML} 2016, New York City, NY, USA, June 19-24, 2016}}
  \emph{(\bibinfo{series}{{JMLR} Workshop and Conference Proceedings},
  Vol.~\bibinfo{volume}{48})}. \bibinfo{publisher}{JMLR.org},
  \bibinfo{pages}{2071--2080}.
\newblock
\urldef\tempurl%
\url{http://proceedings.mlr.press/v48/trouillon16.html}
\showURL{%
\tempurl}


\bibitem[Valiev et~al\mbox{.}(2018)]%
        {fse18pypi}
\bibfield{author}{\bibinfo{person}{Marat Valiev}, \bibinfo{person}{Bogdan
  Vasilescu}, {and} \bibinfo{person}{James~D. Herbsleb}.}
  \bibinfo{year}{2018}\natexlab{}.
\newblock \showarticletitle{Ecosystem-level determinants of sustained activity
  in open-source projects: a case study of the PyPI ecosystem}. In
  \bibinfo{booktitle}{\emph{Proceedings of the 2018 {ACM} Joint Meeting on
  European Software Engineering Conference and Symposium on the Foundations of
  Software Engineering, {ESEC/SIGSOFT} {FSE} 2018, Lake Buena Vista, FL, USA,
  November 04-09, 2018}}. \bibinfo{publisher}{{ACM}},
  \bibinfo{pages}{644--655}.
\newblock
\urldef\tempurl%
\url{https://doi.org/10.1145/3236024.3236062}
\showDOI{\tempurl}


\bibitem[van~der Burg et~al\mbox{.}(2014)]%
        {ase14license}
\bibfield{author}{\bibinfo{person}{Sander van~der Burg}, \bibinfo{person}{Eelco
  Dolstra}, \bibinfo{person}{Shane McIntosh}, \bibinfo{person}{Julius Davies},
  \bibinfo{person}{Daniel~M. Germ{\'{a}}n}, {and} \bibinfo{person}{Armijn
  Hemel}.} \bibinfo{year}{2014}\natexlab{}.
\newblock \showarticletitle{Tracing software build processes to uncover license
  compliance inconsistencies}. In \bibinfo{booktitle}{\emph{{ACM/IEEE}
  International Conference on Automated Software Engineering, {ASE} '14,
  Vasteras, Sweden - September 15 - 19, 2014}}. \bibinfo{publisher}{{ACM}},
  \bibinfo{pages}{731--742}.
\newblock
\urldef\tempurl%
\url{https://doi.org/10.1145/2642937.2643013}
\showDOI{\tempurl}


\bibitem[Wang et~al\mbox{.}(2019)]%
        {fse2019glossary}
\bibfield{author}{\bibinfo{person}{Chong Wang}, \bibinfo{person}{Xin Peng},
  \bibinfo{person}{Mingwei Liu}, \bibinfo{person}{Zhenchang Xing},
  \bibinfo{person}{Xuefang Bai}, \bibinfo{person}{Bing Xie}, {and}
  \bibinfo{person}{Tuo Wang}.} \bibinfo{year}{2019}\natexlab{}.
\newblock \showarticletitle{A learning-based approach for automatic
  construction of domain glossary from source code and documentation}. In
  \bibinfo{booktitle}{\emph{Proceedings of the {ACM} Joint Meeting on European
  Software Engineering Conference and Symposium on the Foundations of Software
  Engineering, {ESEC/SIGSOFT} {FSE} 2019, Tallinn, Estonia, August 26-30,
  2019}}. \bibinfo{publisher}{{ACM}}, \bibinfo{pages}{97--108}.
\newblock
\urldef\tempurl%
\url{https://doi.org/10.1145/3338906.3338963}
\showDOI{\tempurl}


\bibitem[Wang et~al\mbox{.}(2023a)]%
        {TSE23CONCEPTLINK}
\bibfield{author}{\bibinfo{person}{Chong Wang}, \bibinfo{person}{Xin Peng},
  \bibinfo{person}{Zhenchang Xing}, {and} \bibinfo{person}{Xiujie Meng}.}
  \bibinfo{year}{2023}\natexlab{a}.
\newblock \showarticletitle{Beyond Literal Meaning: Uncover and Explain
  Implicit Knowledge in Code Through Wikipedia-Based Concept Linking}.
\newblock \bibinfo{journal}{\emph{{IEEE} Trans. Software Eng.}}
  \bibinfo{volume}{49}, \bibinfo{number}{5} (\bibinfo{year}{2023}),
  \bibinfo{pages}{3226--3240}.
\newblock
\urldef\tempurl%
\url{https://doi.org/10.1109/TSE.2023.3250029}
\showDOI{\tempurl}


\bibitem[Wang et~al\mbox{.}(2023b)]%
        {wang2023xcos}
\bibfield{author}{\bibinfo{person}{Chong Wang}, \bibinfo{person}{Xin Peng},
  \bibinfo{person}{Zhenchang Xing}, \bibinfo{person}{Yue Zhang},
  \bibinfo{person}{Mingwei Liu}, \bibinfo{person}{Rong Luo}, {and}
  \bibinfo{person}{Xiujie Meng}.} \bibinfo{year}{2023}\natexlab{b}.
\newblock \showarticletitle{XCoS: Explainable Code Search based on Query
  Scoping and Knowledge Graph}.
\newblock \bibinfo{journal}{\emph{ACM Transactions on Software Engineering and
  Methodology}} (\bibinfo{year}{2023}).
\newblock


\bibitem[Wang et~al\mbox{.}(2021)]%
        {2021milvus}
\bibfield{author}{\bibinfo{person}{Jianguo Wang}, \bibinfo{person}{Xiaomeng
  Yi}, \bibinfo{person}{Rentong Guo}, \bibinfo{person}{Hai Jin},
  \bibinfo{person}{Peng Xu}, \bibinfo{person}{Shengjun Li},
  \bibinfo{person}{Xiangyu Wang}, \bibinfo{person}{Xiangzhou Guo},
  \bibinfo{person}{Chengming Li}, \bibinfo{person}{Xiaohai Xu},
  \bibinfo{person}{Kun Yu}, \bibinfo{person}{Yuxing Yuan},
  \bibinfo{person}{Yinghao Zou}, \bibinfo{person}{Jiquan Long},
  \bibinfo{person}{Yudong Cai}, \bibinfo{person}{Zhenxiang Li},
  \bibinfo{person}{Zhifeng Zhang}, \bibinfo{person}{Yihua Mo},
  \bibinfo{person}{Jun Gu}, \bibinfo{person}{Ruiyi Jiang}, \bibinfo{person}{Yi
  Wei}, {and} \bibinfo{person}{Charles Xie}.} \bibinfo{year}{2021}\natexlab{}.
\newblock \showarticletitle{Milvus: {A} Purpose-Built Vector Data Management
  System}. In \bibinfo{booktitle}{\emph{{SIGMOD} '21: International Conference
  on Management of Data, Virtual Event, China, June 20-25, 2021}}.
  \bibinfo{publisher}{{ACM}}, \bibinfo{pages}{2614--2627}.
\newblock
\urldef\tempurl%
\url{https://doi.org/10.1145/3448016.3457550}
\showDOI{\tempurl}


\bibitem[Wang et~al\mbox{.}(2017b)]%
        {wang2017construct}
\bibfield{author}{\bibinfo{person}{Lu Wang}, \bibinfo{person}{Xiaobing Sun},
  \bibinfo{person}{Jingwei Wang}, \bibinfo{person}{Yucong Duan}, {and}
  \bibinfo{person}{Bin Li}.} \bibinfo{year}{2017}\natexlab{b}.
\newblock \showarticletitle{Construct bug knowledge graph for bug resolution:
  poster}. In \bibinfo{booktitle}{\emph{Proceedings of the 39th International
  Conference on Software Engineering, {ICSE} 2017, Buenos Aires, Argentina, May
  20-28, 2017 - Companion Volume}}. \bibinfo{publisher}{{IEEE} Computer
  Society}, \bibinfo{pages}{189--191}.
\newblock
\urldef\tempurl%
\url{https://doi.org/10.1109/ICSE-C.2017.102}
\showDOI{\tempurl}


\bibitem[Wang et~al\mbox{.}(2017a)]%
        {kgesurvey}
\bibfield{author}{\bibinfo{person}{Quan Wang}, \bibinfo{person}{Zhendong Mao},
  \bibinfo{person}{Bin Wang}, {and} \bibinfo{person}{Li Guo}.}
  \bibinfo{year}{2017}\natexlab{a}.
\newblock \showarticletitle{Knowledge Graph Embedding: {A} Survey of Approaches
  and Applications}.
\newblock \bibinfo{journal}{\emph{{IEEE} Trans. Knowl. Data Eng.}}
  \bibinfo{volume}{29}, \bibinfo{number}{12} (\bibinfo{year}{2017}),
  \bibinfo{pages}{2724--2743}.
\newblock
\urldef\tempurl%
\url{https://doi.org/10.1109/TKDE.2017.2754499}
\showDOI{\tempurl}


\bibitem[Welch(1947)]%
        {welch1947generalization}
\bibfield{author}{\bibinfo{person}{Bernard~L Welch}.}
  \bibinfo{year}{1947}\natexlab{}.
\newblock \showarticletitle{The generalization of ‘STUDENT'S’problem when
  several different population varlances are involved}.
\newblock \bibinfo{journal}{\emph{Biometrika}} \bibinfo{volume}{34},
  \bibinfo{number}{1-2} (\bibinfo{year}{1947}), \bibinfo{pages}{28--35}.
\newblock


\bibitem[Xie et~al\mbox{.}(2020)]%
        {fse20funcverb}
\bibfield{author}{\bibinfo{person}{Wenkai Xie}, \bibinfo{person}{Xin Peng},
  \bibinfo{person}{Mingwei Liu}, \bibinfo{person}{Christoph Treude},
  \bibinfo{person}{Zhenchang Xing}, \bibinfo{person}{Xiaoxin Zhang}, {and}
  \bibinfo{person}{Wenyun Zhao}.} \bibinfo{year}{2020}\natexlab{}.
\newblock \showarticletitle{{API} method recommendation via explicit matching
  of functionality verb phrases}. In \bibinfo{booktitle}{\emph{28th ACM Joint
  Meeting on European Software Engineering Conference and Symposium on the
  Foundations of Software Engineering, {ESEC/SIGSOFT} {FSE} 2020, November
  8-13, 2020, Virtual Event, USA}}. \bibinfo{publisher}{{ACM}},
  \bibinfo{pages}{1015--1026}.
\newblock
\urldef\tempurl%
\url{https://doi.org/10.1145/3368089.3409731}
\showDOI{\tempurl}


\bibitem[Xing et~al\mbox{.}(2021)]%
        {jos2021automatic}
\bibfield{author}{\bibinfo{person}{Shuangshuang Xing}, \bibinfo{person}{Mingwei
  Liu}, {and} \bibinfo{person}{Xin Peng}.} \bibinfo{year}{2021}\natexlab{}.
\newblock \showarticletitle{Automatic Code Semantic Tag Generation Approach
  Based on Software Knowledge Graph}.
\newblock \bibinfo{journal}{\emph{Journal of Software}} \bibinfo{volume}{33},
  \bibinfo{number}{11} (\bibinfo{year}{2021}), \bibinfo{pages}{4027--4045}.
\newblock


\bibitem[Yang et~al\mbox{.}(2015)]%
        {DistMult}
\bibfield{author}{\bibinfo{person}{Bishan Yang}, \bibinfo{person}{Wen{-}tau
  Yih}, \bibinfo{person}{Xiaodong He}, \bibinfo{person}{Jianfeng Gao}, {and}
  \bibinfo{person}{Li Deng}.} \bibinfo{year}{2015}\natexlab{}.
\newblock \showarticletitle{Embedding Entities and Relations for Learning and
  Inference in Knowledge Bases}. In \bibinfo{booktitle}{\emph{3rd International
  Conference on Learning Representations, {ICLR} 2015, San Diego, CA, USA, May
  7-9, 2015, Conference Track Proceedings}}.
\newblock
\urldef\tempurl%
\url{http://arxiv.org/abs/1412.6575}
\showURL{%
\tempurl}


\bibitem[Zhang et~al\mbox{.}(2020)]%
        {icpc20apimigration}
\bibfield{author}{\bibinfo{person}{Zejun Zhang}, \bibinfo{person}{Minxue Pan},
  \bibinfo{person}{Tian Zhang}, \bibinfo{person}{Xinyu Zhou}, {and}
  \bibinfo{person}{Xuandong Li}.} \bibinfo{year}{2020}\natexlab{}.
\newblock \showarticletitle{Deep-Diving into Documentation to Develop Improved
  Java-to-Swift {API} Mapping}. In \bibinfo{booktitle}{\emph{{ICPC} '20: 28th
  International Conference on Program Comprehension, Seoul, Republic of Korea,
  July 13-15, 2020}}. \bibinfo{publisher}{{ACM}}, \bibinfo{pages}{106--116}.
\newblock
\urldef\tempurl%
\url{https://doi.org/10.1145/3387904.3389282}
\showDOI{\tempurl}


\end{thebibliography}

\end{document}